\documentclass [prl,twocolumn,superscriptaddress]{revtex4-1}

\usepackage{graphicx}
\usepackage{subfigure}
\usepackage{amsmath}
\usepackage{mathrsfs}
\usepackage{amsthm}
\usepackage{amssymb}
\usepackage{hyperref}
\usepackage{xcolor}
\usepackage{csquotes}
\usepackage{color}

\hypersetup{colorlinks=true, linkcolor=black, citecolor=black, urlcolor=blue}
\allowdisplaybreaks

\newcommand{\ket}[1]{\left| #1 \right>}

\begin{document}

%==================================================================================================
\title{Driven-Dissipative Supersolid in a Ring Cavity}
\author{Farokh Mivehvar}
\email[Corresponding author: ]{farokh.mivehvar@uibk.ac.at}
\affiliation{Institut f\"ur Theoretische Physik, Universit{\"a}t Innsbruck, A-6020~Innsbruck, Austria}
\author{Stefan Ostermann}
\affiliation{Institut f\"ur Theoretische Physik, Universit{\"a}t Innsbruck, A-6020~Innsbruck, Austria}
\author{Francesco Piazza}
\affiliation{Max-Planck-Institut f\"{u}r Physik komplexer Systeme, D-01187 Dresden, Germany}
\author{Helmut Ritsch}
\email[Corresponding author: ]{helmut.ritsch@uibk.ac.at}
\affiliation{Institut f\"ur Theoretische Physik, Universit{\"a}t Innsbruck, A-6020~Innsbruck, Austria}

\begin{abstract}
Supersolids are characterized by the counter-intuitive coexistence of
superfluid and crystalline order.
Here we study a supersolid phase emerging in the steady state of a
driven-dissipative system. We consider a transversely pumped 
Bose-Einstein condensate trapped along the axis of a ring cavity and 
coherently coupled to a pair of degenerate counter-propagating cavity modes.
Above a threshold pump strength the interference of photons scattered into the 
two cavity modes results in an emergent superradiant
lattice, which spontaneously breaks the continuous translational symmetry towards a periodic atomic pattern. 
The crystalline steady state inherits the superfluidity of 
the Bose-Einstein condensate, thus exhibiting genuine properties of a supersolid. 
A gapless collective Goldstone mode correspondingly appears in the superradiant phase, 
which can be non-destructively monitored via the relative phase of the two cavity modes on the cavity output. 
Despite cavity-photon losses the Goldstone mode remains undamped,
indicating the robustness of the supersolid phase.
\end{abstract}

\maketitle

%==================================================================================================
\emph{Introduction.}---A supersolid behaves as both a crystalline solid and a superfluid.
It \textit{spontaneously} breaks two continuous symmetries, 
namely the external spatial translation invariance 
and the internal superfluid gauge invariance. 
That is, it simultaneously possesses diagonal (i.e., density) 
and off-diagonal (i.e., superfluid) long-range orders~\cite{Boninsegni2012}. 
This paradoxical state of matter has been predicted almost 50 years ago to exist in solid 
helium-4~\cite{Gross1957,Thouless1969,Andreev1969,Chester1970,Leggett1970}. 
Despite intensive experimental efforts~\cite{Kim2004a,Kim2004b},  
supersolidity has not been conclusively observed in solid 
helium-4 yet~\cite{Day2007,Kim2012NoSS,Maris2012}.  
 
In a different direction, very recently clear signatures of supersolidity have been observed 
in weakly interacting ultracold atomic systems.
At MIT, synthetic spin-orbit coupling was induced in a multi-component Bose-Einstein condensate (BEC)~\cite{Li2017}. 
The ground state of the system spontaneously breaks the continuous translational symmetry and 
forms a density modulated stripe pattern, while maintaining superfluidity of the BEC. At ETH,
a transversally driven BEC was symmetrically coupled to two modes of two crossed linear cavities~\cite{Lonard2017}. 
Interference of pump-laser photons and photons scattered into the cavity modes 
yields an emergent superradiant optical lattice for the BEC, 
which spontaneously breaks the continuous translational invariance 
towards a density-modulated superfluid state. 
In another experiment at MIT~\cite{Dimitrova2017}, 
a BEC illuminated by two non-interfering counter-propagating lasers 
exhibited collective Rayleigh scattering, resulting in 
spontaneous crystallization of both matter and light~\cite{Ostermann2016,ostermann2017probing}. 
However, the potential appearance of supersolidity in this system has not been thoroughly investigated yet. 
%All theses quantum gas supersolids differ from the supersolid 
%state in solid $^4$He by the fact that they are not characterized by 
%the delocalization of zero-point defects~\cite{Andreev1969,Chester1970} 
%but rather enter the weakly-interacting scenario~\cite{PhysRev.106.161}. 

\begin{figure}[b]
\centering
\includegraphics [width=0.48\textwidth]{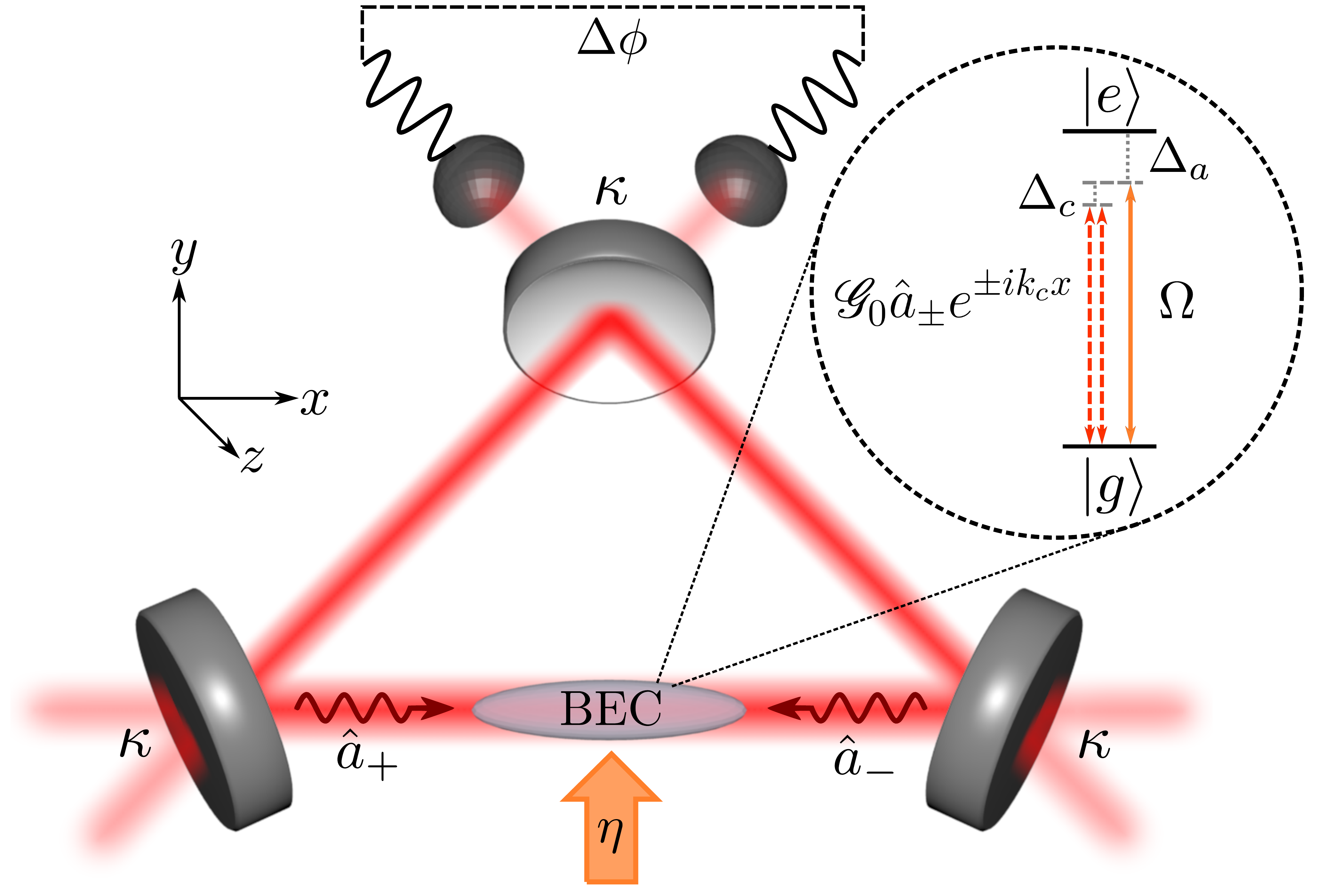}
\caption{Schematic sketch of a BEC inside a ring cavity. 
An internal atomic transition $\ket{g}\leftrightarrow\ket{e}$ is off-resonantly driven 
by a transverse plane-wave laser with Rabi frequency $\Omega$.  
This transition is also off-resonantly coupled to a pair of degenerate, counter-propagating cavity modes $\hat{a}_\pm$ 
with coupling strength  $\mathscr{G}(x)=\mathscr{G}_0 e^{\pm i k_cx}$.}
\label{fig:ring-cavity-geom-coupling}
\end{figure}

Based on the state of the art in experimental quantum-gas cavity QED~\cite{Kruse2003,Nagorny2003,Zimmermann2007,Bux2013,Zimmermann2014,Naik-ringcavitysetup}, 
we propose a novel scheme to experimentally realize and study supersolidity 
in a BEC trapped within a ring resonator~\cite{Moore1990,Horak2001,Nagy2006,Chen2010,Ostermann2015}.  
The BEC, which is transversely illuminated by a standing-wave laser, 
is trapped along the cavity axis in a quasi one-dimensional geometry and dispersively 
coupled to a pair of degenerate counter-propagating field modes as depicted in Fig.~\ref{fig:ring-cavity-geom-coupling}. 
This comprises an intrinsically driven-dissipative system due to the pump laser and cavity-photon losses~\cite{Ritsch2013}.
Therefore, the emergent supersolid above the self-ordering threshold is the steady-state of the system. 
The interesting questions which arise are if and how the supersolid features are modified with respect to thermal
equilibrium.
%This system even has a rich phase diagram
%for a thermal cold gas, at a self-consistent temperature set by the cavity decay rate~\cite{Nagy2006}. 
%The physics is even richer in the zero-temperature limit studied in this work. 

Above a critical laser intensity,
the collective constructive scattering of pump-laser photons into the cavity modes 
results in an emergent superradiant optical lattice. 
In contrast to the standing-wave linear cavity~\cite{Baumann2010}, 
the running-wave ring cavity respects the \textit{continuous} spatial translational symmetry. 
Hence, the location of the emerging optical lattice is not pre-determined 
by the cavity mirrors and \textit{spontaneously} breaks the continuous translational symmetry, 
similar to the emergent optical lattice in the crossed-cavity experiment~\cite{Lonard2017}.  
Nevertheless, in the latter the continuous symmetry is merely an approximate and 
fine-tuned symmetry~\cite{Lang2017,Moodie2017,Morales2017}.
A similar continuous symmetry breaking can also occur 
%in ring resonators with longitudinal pump~\cite{Ostermann2015} or
for atoms trapped close to optical fibers~\cite{Griesser2013,chang2013}.

\begin{figure*}[t]
\centering
\includegraphics [width=0.9\textwidth]{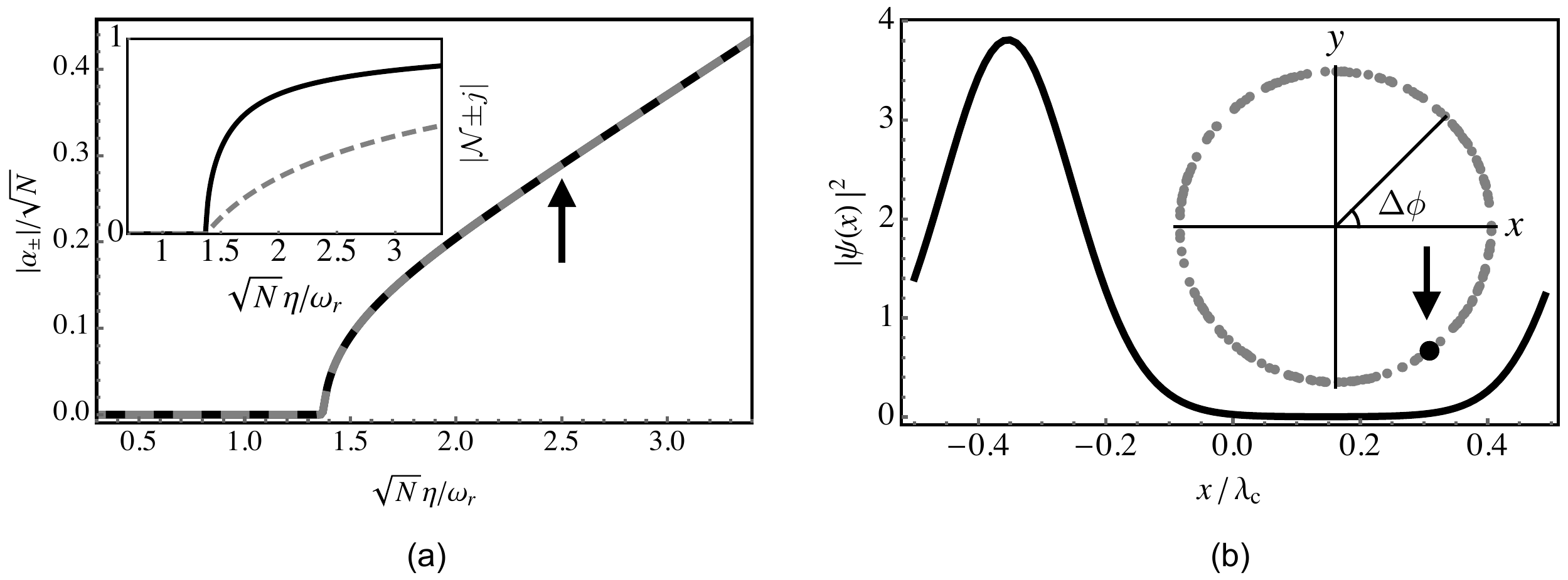}
\caption{
%(Color online) 
Dicke superradiant phase transition and atomic self-organization. (a) The absolute values of 
the rescaled cavity-field amplitudes $|\alpha_\pm|/\sqrt{N}$
(black solid and grey dashed curves, respectively) are shown 
as a function of the rescaled effective cavity-pump strength $\sqrt{N}\eta/\omega_r$.
%The absolute values of the two modes are indistinguishable from one another.
The superradiant phase transition occurs at the critical pump strength $\sqrt{N}\eta_c\approx1.38\omega_r$,
where $|\alpha_+|=|\alpha_-|>0$.
The inset shows the absolute values of the quantities 
$|\mathcal{N}_{\pm 1}|$ (the solid black curve) and 
$|\mathcal{N}_{\pm 2}|$ (the dashed gray curve) as 
a function of $\sqrt{N}\eta/\omega_r$.
(b) A typical self-ordered atomic density profile is shown for $\sqrt{N}\eta=2.5\omega_r$
with $\Delta\phi\approx1.71\pi$ and $\Phi\approx0.09\pi$, where $\Delta\phi$ 
fixes the position of the density maximum $x_m\approx-0.35\lambda_c$.
The inset illustrates the distribution of $\Delta\phi$ 
for 200 numerical runs for $\sqrt{N}\eta=2.5\omega_r$,
exhibiting the continuous $U(1)$ symmetry breaking.
The parameters are set to $(\Delta_c,U,\kappa)=(-8,-1,2)\omega_r$.}
\label{fig:alphas-Rphase-orderP-density}
\end{figure*}

The emergent superradiant lattice drives the BEC into a 
density modulated state --- i.e., a crystalline phase --- with the 
spontaneously broken continuous translational symmetry [see Fig.~\ref{fig:alphas-Rphase-orderP-density}(b)]. 
It, nevertheless, inherits superfluidity of the BEC with a long-range phase coherence. 
Therefore, the resultant steady state in the superradiant phase is a supersolid. 
As the cavity comprises an open system, the continuous symmetry breaking 
can be monitored non-destructively in real time via the cavity output, 
namely via the \textit{relative phase} of the two cavity modes~\cite{Kruse2003,Nagorny2003}. 
In particular, the relative phase takes a random value between $0$ and $2\pi$ in the superradiant phase,
spontaneously breaking the continuous $U(1)$ symmetry 
[see the inset of Fig.~\ref{fig:alphas-Rphase-orderP-density}(b)].  
In fact, at the onset of superradiance a superposition of field amplitudes 
with different phases correlated with density fluctuations 
emerges, forming a highly entangled atom-field state. 
This state subsequently collapses to a state with a certain random relative phase 
via quantum jumps induced by cavity photon 
losses~\cite{maschler2007entanglement,vukics2007microscopic,kramer2014self}.

\begin{figure}[t!]
\centering
\includegraphics [width=0.48\textwidth]{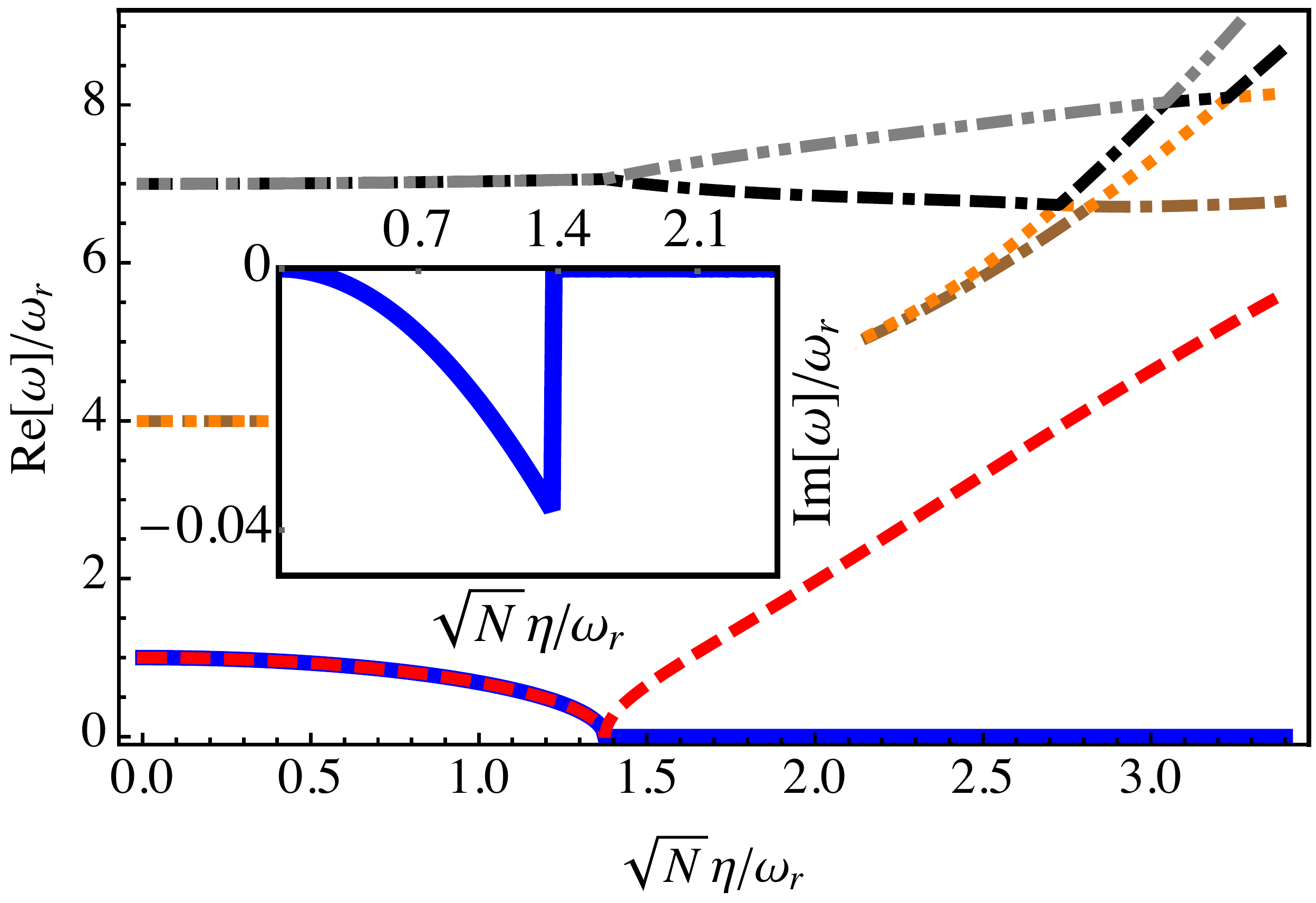}
\caption{
(Color online) 
Low-lying collective excitations.
The real part of the six lowest-lying excitation spectra are shown as 
a function of the rescaled effective cavity-pump strength $\sqrt{N}\eta/\omega_r$.
By increasing $\sqrt{N}\eta/\omega_r$ from zero the lowest excitation, 
corresponding mainly to the atomic condensate fluctuations with momenta $\pm\hbar k_c$,
softens and the excitation gap closes at $\sqrt{N}\eta_{\rm G}\approx1.37\omega_r$.
By further increasing $\sqrt{N}\eta/\omega_r$, a gap does not
open in the first excitation branch (the solid blue curve), indicating that this is a gapless
Goldstone mode corresponding to the spontaneously
broken continuous $U(1)$ symmetry. The inset shows the imaginary
part of the Goldstone mode. 
The parameters are the same as Fig.~\ref{fig:alphas-Rphase-orderP-density}.
}
\label{fig:excitations}
\end{figure}

Analysis of collective excitations confirms the supersolidity of the superradiant steady state.
At the onset of the superradiant phase transition, where the continuous
$U(1)$ symmetry is spontaneously broken, a gapless Goldstone mode appears in
the spectrum of collective excitations. Unlike all other collective modes, the
Goldstone mode remains undamped despite cavity losses (see
Fig.~\ref{fig:excitations}). This is due to the fact that photon losses do not
affect the relative phase and preserve the $U(1)$ symmetry. 
This is in contrast to the supersolid realized in the crossed-cavity
setup, where the origin of the $U(1)$ symmetry is different and associated with 
the freedom of the photon redistribution between the two cavity modes~\cite{Lonard2017}.
Therefore, photon losses do not respect the $U(1)$ symmetry and 
should result in damping of the Goldstone mode~\cite{leonard2017monitoring}.

%==================================================================================================
\emph{Model.}---Consider bosonic two-level atoms trapped along the axis 
of a ring resonator by a tight confining potential along the transverse directions. 
The atoms are illuminated from the side by an off-resonant, standing-wave external pump 
laser as depicted in Fig.~\ref{fig:ring-cavity-geom-coupling}, which induces the transition 
$\ket{g}\leftrightarrow\ket{e}$ with the Rabi frequency $\Omega$. Furthermore, 
the transition $\ket{g}\leftrightarrow\ket{e}$ is also off-resonantly coupled to a pair of degenerate, counter-propagating 
cavity modes $\hat{a}_\pm$, with coupling strength 
$\mathscr{G}(x)=\mathscr{G}_0 e^{\pm i k_cx}$. 
The cavity modes are initially in the vacuum state.
The pump and cavity frequencies, respectively, 
$\omega_p$ and $\omega_c=ck_c$ are assumed to be near resonant with each other, but far-red detuned 
with respect to the atomic frequency $\omega_a$.

In the dispersive regime $|\Delta_a|\equiv|\omega_p-\omega_a|\gg\{\Omega,\mathscr{G}_0\}$, 
the atomic excited state $\ket{e}$ reaches
quickly to a steady state  with a negligible population and its dynamics can be adiabatically eliminated~\cite{Ritsch2013}.
This yields an effective Hamiltonian for the atomic ground state and the cavity modes,
$H_{\rm eff}=
\int \hat\psi^\dag(x)\mathcal{H}_{\rm eff}^{(1)}\hat\psi(x)dx
-\hbar\Delta_c(\hat{a}_+^\dagger\hat{a}_++\hat{a}_-^\dagger\hat{a}_-),$ 
with the effective single-particle atomic Hamiltonian density:
\begin{align} \label{eq:single-body-eff-H}
\mathcal{H}_{\rm eff}^{(1)}&=
-\frac{\hbar^2}{2m}\frac{\partial^2}{\partial x^2}
+\hbar U \Big(\hat{a}_+^\dag\hat{a}_++\hat{a}_-^\dag\hat{a}_-
+\hat{a}_+^\dag\hat{a}_- e^{-2ik_cx}\nonumber\\
&+\hat{a}_-^\dag\hat{a}_+ e^{2ik_cx}\Big)
+\hbar\eta \Big(\hat{a}_+ e^{ik_cx}+\hat{a}_- e^{-ik_cx}
+\text{H.c.}
\Big).
\end{align}
Here, $\hat\psi(x)$ is the bosonic annihilation field operator for the atomic ground state.
%satisfying the canonical bosonic commutation relation 
%$[\hat\psi(x),\hat\psi^\dag(x')]=\delta(x-x')$. 
We have introduced the cavity detuning with respect to the pump $\Delta_c\equiv\omega_p-\omega_c$,
the maximum depth of the optical potential per photon due to two-photon scattering between cavity modes 
$\hbar U\equiv\hbar\mathscr{G}_0^2/\Delta_a$  and 
the maximum depth of the optical potential per photon due to the two-photon scattering between pump and cavity modes
(or the effective cavity-pump strength) 
$\hbar\eta\equiv\hbar\mathscr{G}_0\Omega/\Delta_a$.
Although finite atom-atom interactions are needed to ensure the superfluidity of the BEC, we have assumed them to be negligibly
small with respect to the cavity-mediated interactions. This is quantitatively a good approximation for typical
cavity-QED experiments, including the recent observation of the supersolid~\cite{Lonard2017}. 
Finally, the cavity-photon losses with rate $\kappa$ are included via 
Lindblad operators in the master equation for the density matrix $\rho$:
$
\mathcal{L} \rho= \kappa\sum_{\ell=+,-}
\left(2 \hat{a}_\ell\rho \hat{a}_\ell^\dagger - \left\{\hat{a}_\ell^\dagger \hat{a}_\ell, \rho \right\}\right).
$

The system possesses a continuous $U(1)$ symmetry, as
the effective Hamiltonian $H_{\rm eff}$ and the Lindblad operators 
are invariant under the simultaneous spatial translation
$x\to\mathcal{T}_{X}x=x+X$ and cavity-phase rotations
$\hat{a}_\pm\to\mathcal{U}_X\hat{a}_\pm=\hat{a}_\pm e^{\mp i k_c X}$. 
This $U(1)$ symmetry is spontaneously broken in the superradiant phase, as illustrated in the inset of 
Fig.~\ref{fig:alphas-Rphase-orderP-density}(b), 
where $\langle\hat{a}_\pm\rangle$ acquire non-zero values with arbitrary phases.
%The BEC density is then modulated with maxima located at positions 
%fixed by the relative cavity phase as depicted in Fig.~\ref{fig:alphas-Rphase-orderP-density}(b).
%Corresponding to this continuous $U(1)$ symmetry breaking, a gapless Goldstone mode appears 
%in the spectrum of elementary excitations, as shown in Fig.~\ref{fig:excitations}. 

%==================================================================================================
%-------FIGURE--------------- 

\emph{Mean-Field Approach and Continuous Symmetry Breaking.}---In the thermodynamic limit,
where the mean-field approximation becomes accurate~\cite{piazza2013bose},
the system is described by a set of three coupled 
mean-field (Heisenberg) equations for the cavity-field amplitudes 
$\langle\hat{a}_\pm(t)\rangle=\alpha_\pm(t)=|\alpha_\pm(t)|e^{i\phi_\pm(t)}$  
and the atomic condensate wavefunction 
$\langle \hat\psi(x,t)\rangle=\psi(x,t)=\sqrt{n(x,t)}e^{i\theta(t)}$~\cite{SM-1BEC-2ringcavity},
\begin{align} \label{eq:MF-Heisenberg-eqs}
i\frac{\partial}{\partial t}\alpha_\pm&=
\left(-\Delta_c+UN-i\kappa\right)\alpha_\pm
+U\mathcal{N}_{\pm2}\alpha_\mp
+\eta\mathcal{N}_{\pm1},
\nonumber\\
i\hbar\frac{\partial}{\partial t}\psi&=\mathcal{H}_{\rm eff}^{(1)}\psi,
\end{align}
where $N=\int n(x) dx$ is the number of the particles. 
%Here, we have added field damping terms proportional to the cavity  decay rate $\kappa$. 
One can identify $\mathcal{N}_{\pm1}\equiv\int n(x)e^{\mp ik_cx}dx$
as the atomic order parameters, dual to the cavity order
parameters $\alpha_{\pm}$, which characterize the 
probability of the photon scattering between the pump and cavity modes with $\mp\hbar k_c$ momentum
transfer to the atoms along the cavity axis $x$. Whereas, $\mathcal{N}_{\pm2}\equiv\int n(x)e^{\mp2ik_cx}dx$
quantifies the probability of the photon redistribution between the two cavity modes with $\mp2\hbar k_c$ momentum
transfer to the atoms in the $x$ direction.

We self-consistently find the steady-state solutions of Eq.~\eqref{eq:MF-Heisenberg-eqs} 
by setting $\partial_t\alpha_\pm=0$ and $i\hbar\partial_t\psi=\mu\psi$, with $\mu$ being the chemical potential.
Figure~\ref{fig:alphas-Rphase-orderP-density}(a) shows the absolute values 
of the rescaled cavity-mode amplitudes $|\alpha_\pm|/\sqrt{N}$ 
(black solid and grey dashed curves, respectively) 
as a function of the rescaled effective pump strength $\sqrt{N}\eta/\omega_r$, 
with $\omega_r\equiv\hbar k_c^2/2m$ being the recoil frequency. 
Below the threshold pump strength $\sqrt{N}\eta_c\approx1.38\omega_r$, 
the cavity modes are empty and the BEC is uniform. 
By increasing the pump strength above $\eta_c$ the system undergoes a
superradiant phase transition, where the cavity amplitudes acquire non-zero values
$|\alpha_+|=|\alpha_-|=|\alpha|$.
%The parameters are set to $\Delta_c=-8\omega_r$, $U=-\omega_r$, and $\kappa=2\omega_r$. 
  
In the superradiant state, the relative phase
$\Delta\phi\equiv(\phi_+-\phi_-)/2$ of the two cavity modes is fixed
in an arbitrary value between $0$ and $2\pi$
and the continuous $U(1)$ symmetry is, therefore, spontaneously broken.
This is illustrated in 
the inset of Fig.~\ref{fig:alphas-Rphase-orderP-density}(b), where the distribution of $\Delta\phi$ 
is shown for 200 numerical runs for a pump strength $\sqrt{N}\eta=2.5\omega_r$ 
[indicated by the arrow in Fig.~\ref{fig:alphas-Rphase-orderP-density}(a)].
%In the Supplemental Material, we also present arguments and numerical examples 
%for this spontaneous symmetry breaking 
%in the single-particle limit~\cite{SM-1BEC-2ringcavity}.  
%While the relative phase $\Delta\phi$ in each run of the numerical simulation is randomly fixed, 
%the total phase $\phi_++\phi_-$ is always locked at zero (modulo $2\pi$). 
The emergent superradiant lattice has the form
$V_{\rm SR}(x)=2U|\alpha|^2
\cos(2k_cx+2\Delta\phi)+4\eta|\alpha|\cos(k_cx+\Delta\phi)\cos(\Phi)$, 
with $\Phi\equiv(\phi_++\phi_-)/2$ being the total phase. The spontaneously chosen value of
$\Delta\phi$ fixes the position of the lattice minima and thus of the BEC
density modulation, spontaneously breaking the continuous translational invariance
and resulting in a supersolid state. 
A typical self-ordered, $\lambda_c$-periodic atomic density profile is shown in
Fig.~\ref{fig:alphas-Rphase-orderP-density}(b) for $\sqrt{N}\eta=2.5\omega_r$ 
with $\Delta\phi\approx1.71\pi$ 
[the pronounced black dot indicated by the arrow 
in the inset of Fig.~\ref{fig:alphas-Rphase-orderP-density}(b)]
and $\Phi\approx0.09\pi$.

The total phase $\Phi$ solely modifies the lattice amplitude, except the special case of
$\Phi=\pi/2$ where the lattice spacing is reduced from $\lambda_c$
to $\lambda_c/2$. Note that $\Phi$ is not random as it is invariant under the 
$U(1)$ phase rotation $\alpha_\pm\to\alpha_\pm e^{\mp ik_cX}$. 
The total phase $\Phi$ solely depends on $\kappa$, while
the relative phase $\Delta\phi$ is independent of $\kappa$.
This is becuasse photon losses induce 
equal extra phase shifts for both cavity-field amplitudes.
%Since photon losses (i.e., $\kappa\neq0$) induce 
%equal extra phase shifts for both cavity-field amplitudes, 
%only the total phase $\Phi$ depends on $\kappa$.
%while the relative phase $\Delta\phi$ is independent of it. 
Therefore, the spontaneous $U(1)$-symmetry breaking and thus the supersolid order
persist even in the presence of dissipation. 

%Note that a non-zero cavity-field decay rate $\kappa\neq0$ induces extra constant
%phase factors for the cavity-field amplitudes~\cite{Baumann2011}. However,
%this would not affect the physics discussed here.

%==================================================================================================
\emph{Collective Excitations and the Goldstone Mode}.---Let us now turn our attention 
to elementary excitations of the system, which include quantum fluctuations of both condensate wavefunction 
$\delta\psi(x,t)=\delta\psi^{(+)}(x)e^{-i\omega t}
+[\delta\psi^{(-)}(x)]^*e^{i\omega^* t}$
and cavity-field amplitudes $\delta\alpha_\pm(t)=\delta\alpha_\pm^{(+)}e^{-i\omega t}
+[\delta\alpha_\pm^{(-)}]^*e^{i\omega^* t}$ above the mean-field solutions
$\psi_0(x)$ and $\alpha_{0\pm}$ (with the corresponding chemical potential $\mu_0$).
Linearizing Eq.~\eqref{eq:MF-Heisenberg-eqs} yields Bogoliubov-type equations
for the quantum fluctuations ~\cite{Horak2001,Nagy2008,Mivehvar2017b},
\begin{align} \label{eq:linearized-eqs}
i\frac{\partial}{\partial t}\delta\alpha_\pm&=
\left(-\Delta_c+UN-i\kappa\right)\delta\alpha_\pm
+U\mathcal{N}_{\pm2}^{(0)}\delta\alpha_\mp
\nonumber\\
&+\int A_{\pm}(\psi_0^*\delta\psi+\psi_0\delta\psi^*)dx,
\nonumber\\
i\frac{\partial}{\partial t}\delta\psi&=
\frac{1}{\hbar}\left(\mathcal{H}_{\rm eff}^{(1)}-\mu_0\right)\delta\psi
\nonumber\\
&+\psi_0 
\left(A_+^*\delta\alpha_+ + A_-^*\delta\alpha_- +  \text{H.c.}\right),
\end{align}
where $\mathcal{N}_{\pm2}^{(0)}=\int n_0(x)e^{\mp2ik_cx}dx$ and we have defined 
$A_{\pm}(x)\equiv U \left(\alpha_{0\pm}+\alpha_{0\mp} e^{\mp2ik_cx}\right)+\eta e^{\mp ik_cx}$ for shorthands. 
The Bogoliubov equations~\eqref{eq:linearized-eqs} can be recast in a matrix form,
\begin{align} \label{eq:Bog-eq}
\omega\mathbf{f}=\mathbf{M}_{\rm B}\mathbf{f},
\end{align}
where
$\mathbf{f}=(
\delta\alpha_+^{(+)},\delta\alpha_+^{(-)},
\delta\alpha_-^{(+)},\delta\alpha_-^{(-)},
\delta\psi^{(+)},\delta\psi^{(-)})^\mathsf{T}$
and $\mathbf{M}_{\rm B}$  is a non-Hermitian matrix;
%which is a function of the mean-field stationary-state solutions
%as well as the system parameters; 
see the Supplemental Material for the details~\cite{SM-1BEC-2ringcavity}.
The eigenvalues $\omega$ of the Bogoliubov equations~\eqref{eq:Bog-eq}
yield collective excitation spectrum of the system.
We numerically solve Eq.~\eqref{eq:Bog-eq} in one unit cell (of length $\lambda_c$)
with periodic boundary conditions to obtain the collective excitations $\omega$. 

Figure~\ref{fig:excitations} shows the real part of the six lowest-lying excitation frequencies as a function
of the effective cavity-pump strength $\sqrt{N}\eta/\omega_r$. 
At small pump strengths, the excitation spectra are weakly dependent on $\eta$ 
and each branch is doubly degenerate.
The lowest four collective excitations
at frequencies $\sim\omega_r$ (solid blue and dashed red curves) and 
$\sim4\omega_r$ (dotted orange and dashed-dotted brown curves) correspond
to mainly atomic fluctuations with momenta $\pm\hbar k_c$ and $\pm2\hbar k_c$, respectively.
The highest two modes (dashed-dashed-dotted black and dotted-dotted-dashed gray curves) 
at frequencies $\sim-\Delta_c+UN=7\omega_r$ are mostly photon-like fluctuations. 

By increasing $\eta$ the collective modes are increasingly mixed with each other
and begin to split up. In particular, the lowest excitation softens and
the excitation gap closes at the pump strength $\sqrt{N}\eta_{\rm G}\approx1.37\omega_r$. 
By increasing pump strength beyond $\eta_{\rm G}$, 
the lowest excitation splits into two branches. The 
lower one (solid blue curve) remains pinned at zero energy, signaling that it is a 
\textit{gapless} Goldstone mode corresponding to the spontaneously broken continuous $U(1)$ symmetry. 
The gapped branch instead corresponds to a Higgs amplitude mode.
These are reminiscent of the recently observed Goldstone and Higgs modes 
in the crossed-cavity experiment~\cite{leonard2017monitoring}. 
The Goldstone mode in the crossed-cavity experiment, however, should have a small
gap of a few $\omega_r$ due to the fact that the continuous $U(1)$ symmetry
is an approximate symmetry~\cite{Lang2017}.
Note that these are in sharp contrast to the self-organization in a linear cavity,
where only a discrete $\mathbf{Z}_2$ symmetry is spontaneously broken and
the first excitation gap closes at the critical pump strength
but then re-opens again~\cite{Horak2001,Nagy2008,Baumann2011}.

Due to a nonzero cavity-field decay rate $\kappa\neq0$ the excitation frequencies can acquire imaginary parts,
which would indicate the damping of the excitations~\cite{Horak2001,Nagy2008}. 
This would in turn result in friction forces on the atoms.
Above the critical pump strength $\eta_c$, all the collective excitations 
except the gapless Goldstone mode acquire imaginary parts.
This is illustrated in the inset of Fig.~\ref{fig:excitations}, 
which shows the imaginary part of the lowest mode
as a function of $\sqrt{N}\eta/\omega_r$.
Although it is damped for small pump strengths, it vanishes
at the critical pump strength $\sqrt{N}\eta_c\approx1.38\omega_r$ 
(recall that the corresponding real part vanishes at the slightly lower pump strength $\eta_{\rm G}$, where
the damping reaches its maximum value), 
in agreement with the mean-field results [see Fig.~\ref{fig:alphas-Rphase-orderP-density}(a)]. 
This means that the center of mass of the entire modulated BEC can move freely
along the cavity axis without
experiencing any friction, once again illustrating the supersolidity of the system. 
The fact that supersolidity survives even in presence of
dissipation is due to the fact that the corresponding Lindblad
operators respect the $U(1)$-symmetry of the
system. This is in contrast to the
supersolid realized in the crossed-cavity setup, where the Goldstone
mode involves photon-number redistribution between the two cavities and
should therefore be damped by photon losses.

%-------FIGURE--------------- 

Around the critical point $\eta_c$, the atomic momentum states $\pm\hbar k_c$
are the dominant atomic fluctuations coupled to the cavity fluctuations. This
can be seen from the inset of Fig.~\ref{fig:alphas-Rphase-orderP-density}(a), where the
quantities $|\mathcal{N}_{\pm j}|$ with $j=1,2$ are shown as a function of $\sqrt{N}\eta/\omega_r$. 
For $\eta\gtrsim\eta_c$, 
 $\mathcal{N}_{\pm1}$ (the black solid curve) are the dominant quantities.
It is, therefore, a good approximation to restrict atomic fluctuations
to the momentum states $\pm\hbar k_c$.
Using the homogeneous solution (i.e., the solution below the Dicke transition) 
$\alpha_{0\pm}=0$ and $\psi_0=\sqrt{N/\lambda_c}$, 
one can diagonalize Eq.~\eqref{eq:Bog-eq} in this restricted subspace. 
The zero frequency $\omega=0$ solution yields the critical pump strength~\cite{SM-1BEC-2ringcavity},
\begin{align}
\sqrt{N}\eta_c=\sqrt{\frac{(-\Delta_c+UN)^2+\kappa^2}{4(-\Delta_c+UN)}}\sqrt{\omega_r}\approx1.38\omega_r,
\end{align}
which is in full agreement with the numerical results.

\emph{Experimental Detection of the Supersolid State.}---As discussed earlier, 
in our system the spontaneous breaking of the continuous translational symmetry
corresponds to fixing the value of the relative phase of the two cavity modes. This
can be monitored non-destructively by recombining the cavity outputs through a beam spitter. 
In particular, the system can be in real time
repeatedly driven across the superradiant phase transition by 
sweeping the pump strength across the threshold to verify the uniform distribution of the relative phase in the interval $0-2\pi$, 
similar to experiments with linear cavities~\cite{Baumann2011,Lonard2017}. 
Experimental setups coupling a BEC into fields of a ring cavity already exist  
for almost a decade now~\cite{Zimmermann2007,Bux2013,Zimmermann2014,Naik-ringcavitysetup}. Therefore,
the discussed phenomena could be observed with only minimal changes to current state-of-the-art experiments.

%=======================================================================================
\emph{Outlook.}---Our driven-dissipative supersolid is essentially different than other proposed driven-dissipative supersolid
states in Jaynes-Cummings-Hubbard lattices~\cite{Jiasen2013,Bujnowski2014}, in that the latter ones are only the lattice
supersolid with a broken discrete symmetry and no gapless Goldstone mode~\cite{Landig2016}.
Crucially due to the genuine supersolidity and existence of the undamped gapless Goldstone mode, 
our proposal may have applications in precision measurements. 
By monitoring the relative phase between the cavity modes one can non-destructively 
follow the displacement of the BEC in real time. 
However, there is no back-action of the light field onto the BEC motion (apart from the one induced by
the measurement of the phase) due to the existence of the undamped gapless Goldstone mode. 
Therefore, it could be used as a free-falling zero temperature mass for gravitational acceleration measurements, as an alternative to the atomic 
fountains~\cite{berman1997atom,cronin_2009,tinoetal_2010,mueller_atomint_cavity_2015}.
We defer the investigation of the performance of such a device to a future work.

FM is grateful to Tobias Donner and
Manuele Landini for fruitful discussions. 
%FM and HR are supported by the Austrian Science Fund project I1697-N27.
We acknowledge support by the Austrian Science Fund FWF through the projects SFB FoQuS P13 and I1697-N27.

%---------------------------------------------------------------------%%%-------------------------------------------------------------------
%\bibliographystyle{aip}
\bibliography{1BEC-2RCM}

\begin{thebibliography}{50}%
\makeatletter
\providecommand \@ifxundefined [1]{%
 \@ifx{#1\undefined}
}%
\providecommand \@ifnum [1]{%
 \ifnum #1\expandafter \@firstoftwo
 \else \expandafter \@secondoftwo
 \fi
}%
\providecommand \@ifx [1]{%
 \ifx #1\expandafter \@firstoftwo
 \else \expandafter \@secondoftwo
 \fi
}%
\providecommand \natexlab [1]{#1}%
\providecommand \enquote  [1]{``#1''}%
\providecommand \bibnamefont  [1]{#1}%
\providecommand \bibfnamefont [1]{#1}%
\providecommand \citenamefont [1]{#1}%
\providecommand \href@noop [0]{\@secondoftwo}%
\providecommand \href [0]{\begingroup \@sanitize@url \@href}%
\providecommand \@href[1]{\@@startlink{#1}\@@href}%
\providecommand \@@href[1]{\endgroup#1\@@endlink}%
\providecommand \@sanitize@url [0]{\catcode `\\12\catcode `\$12\catcode
  `\&12\catcode `\#12\catcode `\^12\catcode `\_12\catcode `\%12\relax}%
\providecommand \@@startlink[1]{}%
\providecommand \@@endlink[0]{}%
\providecommand \url  [0]{\begingroup\@sanitize@url \@url }%
\providecommand \@url [1]{\endgroup\@href {#1}{\urlprefix }}%
\providecommand \urlprefix  [0]{URL }%
\providecommand \Eprint [0]{\href }%
\providecommand \doibase [0]{http://dx.doi.org/}%
\providecommand \selectlanguage [0]{\@gobble}%
\providecommand \bibinfo  [0]{\@secondoftwo}%
\providecommand \bibfield  [0]{\@secondoftwo}%
\providecommand \translation [1]{[#1]}%
\providecommand \BibitemOpen [0]{}%
\providecommand \bibitemStop [0]{}%
\providecommand \bibitemNoStop [0]{.\EOS\space}%
\providecommand \EOS [0]{\spacefactor3000\relax}%
\providecommand \BibitemShut  [1]{\csname bibitem#1\endcsname}%
\let\auto@bib@innerbib\@empty
%</preamble>
\bibitem [{\citenamefont {Boninsegni}\ and\ \citenamefont
  {Prokof'ev}(2012)}]{Boninsegni2012}%
  \BibitemOpen
  \bibfield  {author} {\bibinfo {author} {\bibfnamefont {M.}~\bibnamefont
  {Boninsegni}}\ and\ \bibinfo {author} {\bibfnamefont {N.~V.}\ \bibnamefont
  {Prokof'ev}},\ }\href {\doibase 10.1103/RevModPhys.84.759} {\bibfield
  {journal} {\bibinfo  {journal} {Rev. Mod. Phys.}\ }\textbf {\bibinfo {volume}
  {84}},\ \bibinfo {pages} {759} (\bibinfo {year} {2012})}\BibitemShut
  {NoStop}%
\bibitem [{\citenamefont {Gross}(1957)}]{Gross1957}%
  \BibitemOpen
  \bibfield  {author} {\bibinfo {author} {\bibfnamefont {E.~P.}\ \bibnamefont
  {Gross}},\ }\href {\doibase 10.1103/PhysRev.106.161} {\bibfield  {journal}
  {\bibinfo  {journal} {Phys. Rev.}\ }\textbf {\bibinfo {volume} {106}},\
  \bibinfo {pages} {161} (\bibinfo {year} {1957})}\BibitemShut {NoStop}%
\bibitem [{\citenamefont {Thouless}(1969)}]{Thouless1969}%
  \BibitemOpen
  \bibfield  {author} {\bibinfo {author} {\bibfnamefont {D.}~\bibnamefont
  {Thouless}},\ }\href {\doibase 10.1016/0003-4916(69)90286-3} {\bibfield
  {journal} {\bibinfo  {journal} {Ann. Phys. (N. Y.)}\ }\textbf {\bibinfo
  {volume} {52}},\ \bibinfo {pages} {403} (\bibinfo {year} {1969})}\BibitemShut
  {NoStop}%
\bibitem [{\citenamefont {Andreev}\ and\ \citenamefont
  {Lifshitz}(1969)}]{Andreev1969}%
  \BibitemOpen
  \bibfield  {author} {\bibinfo {author} {\bibfnamefont {A.~F.}\ \bibnamefont
  {Andreev}}\ and\ \bibinfo {author} {\bibfnamefont {I.~M.}\ \bibnamefont
  {Lifshitz}},\ }\href@noop {} {\bibfield  {journal} {\bibinfo  {journal} {Sov.
  Phys. JETP}\ }\textbf {\bibinfo {volume} {29}},\ \bibinfo {pages} {1107}
  (\bibinfo {year} {1969})}\BibitemShut {NoStop}%
\bibitem [{\citenamefont {Chester}(1970)}]{Chester1970}%
  \BibitemOpen
  \bibfield  {author} {\bibinfo {author} {\bibfnamefont {G.~V.}\ \bibnamefont
  {Chester}},\ }\href {\doibase 10.1103/PhysRevA.2.256} {\bibfield  {journal}
  {\bibinfo  {journal} {Phys. Rev. A}\ }\textbf {\bibinfo {volume} {2}},\
  \bibinfo {pages} {256} (\bibinfo {year} {1970})}\BibitemShut {NoStop}%
\bibitem [{\citenamefont {Leggett}(1970)}]{Leggett1970}%
  \BibitemOpen
  \bibfield  {author} {\bibinfo {author} {\bibfnamefont {A.~J.}\ \bibnamefont
  {Leggett}},\ }\href {\doibase 10.1103/PhysRevLett.25.1543} {\bibfield
  {journal} {\bibinfo  {journal} {Phys. Rev. Lett.}\ }\textbf {\bibinfo
  {volume} {25}},\ \bibinfo {pages} {1543} (\bibinfo {year}
  {1970})}\BibitemShut {NoStop}%
\bibitem [{\citenamefont {Kim}\ and\ \citenamefont
  {Chan}(2004{\natexlab{a}})}]{Kim2004a}%
  \BibitemOpen
  \bibfield  {author} {\bibinfo {author} {\bibfnamefont {E.}~\bibnamefont
  {Kim}}\ and\ \bibinfo {author} {\bibfnamefont {M.~H.~W.}\ \bibnamefont
  {Chan}},\ }\href {\doibase 10.1038/nature02220} {\bibfield  {journal}
  {\bibinfo  {journal} {Nature}\ }\textbf {\bibinfo {volume} {427}},\ \bibinfo
  {pages} {225} (\bibinfo {year} {2004}{\natexlab{a}})}\BibitemShut {NoStop}%
\bibitem [{\citenamefont {Kim}\ and\ \citenamefont
  {Chan}(2004{\natexlab{b}})}]{Kim2004b}%
  \BibitemOpen
  \bibfield  {author} {\bibinfo {author} {\bibfnamefont {E.}~\bibnamefont
  {Kim}}\ and\ \bibinfo {author} {\bibfnamefont {M.~H.~W.}\ \bibnamefont
  {Chan}},\ }\href {\doibase 10.1126/science.1101501} {\bibfield  {journal}
  {\bibinfo  {journal} {Science}\ }\textbf {\bibinfo {volume} {305}},\ \bibinfo
  {pages} {1941} (\bibinfo {year} {2004}{\natexlab{b}})}\BibitemShut {NoStop}%
\bibitem [{\citenamefont {Day}\ and\ \citenamefont {Beamish}(2007)}]{Day2007}%
  \BibitemOpen
  \bibfield  {author} {\bibinfo {author} {\bibfnamefont {J.}~\bibnamefont
  {Day}}\ and\ \bibinfo {author} {\bibfnamefont {J.}~\bibnamefont {Beamish}},\
  }\href {\doibase 10.1038/nature06383} {\bibfield  {journal} {\bibinfo
  {journal} {Nature}\ }\textbf {\bibinfo {volume} {450}},\ \bibinfo {pages}
  {853} (\bibinfo {year} {2007})}\BibitemShut {NoStop}%
\bibitem [{\citenamefont {Kim}\ and\ \citenamefont {Chan}(2012)}]{Kim2012NoSS}%
  \BibitemOpen
  \bibfield  {author} {\bibinfo {author} {\bibfnamefont {D.~Y.}\ \bibnamefont
  {Kim}}\ and\ \bibinfo {author} {\bibfnamefont {M.~H.~W.}\ \bibnamefont
  {Chan}},\ }\href {\doibase 10.1103/PhysRevLett.109.155301} {\bibfield
  {journal} {\bibinfo  {journal} {Phys. Rev. Lett.}\ }\textbf {\bibinfo
  {volume} {109}},\ \bibinfo {pages} {155301} (\bibinfo {year}
  {2012})}\BibitemShut {NoStop}%
\bibitem [{\citenamefont {Maris}(2012)}]{Maris2012}%
  \BibitemOpen
  \bibfield  {author} {\bibinfo {author} {\bibfnamefont {H.~J.}\ \bibnamefont
  {Maris}},\ }\href {\doibase 10.1103/PhysRevB.86.020502} {\bibfield  {journal}
  {\bibinfo  {journal} {Phys. Rev. B}\ }\textbf {\bibinfo {volume} {86}},\
  \bibinfo {pages} {020502} (\bibinfo {year} {2012})}\BibitemShut {NoStop}%
\bibitem [{\citenamefont {Li}\ \emph {et~al.}(2017)\citenamefont {Li},
  \citenamefont {Lee}, \citenamefont {Huang}, \citenamefont {Burchesky},
  \citenamefont {Shteynas}, \citenamefont {Top}, \citenamefont {Jamison},\ and\
  \citenamefont {Ketterle}}]{Li2017}%
  \BibitemOpen
  \bibfield  {author} {\bibinfo {author} {\bibfnamefont {J.-R.}\ \bibnamefont
  {Li}}, \bibinfo {author} {\bibfnamefont {J.}~\bibnamefont {Lee}}, \bibinfo
  {author} {\bibfnamefont {W.}~\bibnamefont {Huang}}, \bibinfo {author}
  {\bibfnamefont {S.}~\bibnamefont {Burchesky}}, \bibinfo {author}
  {\bibfnamefont {B.}~\bibnamefont {Shteynas}}, \bibinfo {author}
  {\bibfnamefont {F.~{\c{C}}.}\ \bibnamefont {Top}}, \bibinfo {author}
  {\bibfnamefont {A.~O.}\ \bibnamefont {Jamison}}, \ and\ \bibinfo {author}
  {\bibfnamefont {W.}~\bibnamefont {Ketterle}},\ }\href {\doibase
  10.1038/nature21431} {\bibfield  {journal} {\bibinfo  {journal} {Nature}\
  }\textbf {\bibinfo {volume} {543}},\ \bibinfo {pages} {91} (\bibinfo {year}
  {2017})}\BibitemShut {NoStop}%
\bibitem [{\citenamefont {L{\'{e}}onard}\ \emph
  {et~al.}(2017{\natexlab{a}})\citenamefont {L{\'{e}}onard}, \citenamefont
  {Morales}, \citenamefont {Zupancic}, \citenamefont {Esslinger},\ and\
  \citenamefont {Donner}}]{Lonard2017}%
  \BibitemOpen
  \bibfield  {author} {\bibinfo {author} {\bibfnamefont {J.}~\bibnamefont
  {L{\'{e}}onard}}, \bibinfo {author} {\bibfnamefont {A.}~\bibnamefont
  {Morales}}, \bibinfo {author} {\bibfnamefont {P.}~\bibnamefont {Zupancic}},
  \bibinfo {author} {\bibfnamefont {T.}~\bibnamefont {Esslinger}}, \ and\
  \bibinfo {author} {\bibfnamefont {T.}~\bibnamefont {Donner}},\ }\href
  {\doibase 10.1038/nature21067} {\bibfield  {journal} {\bibinfo  {journal}
  {Nature}\ }\textbf {\bibinfo {volume} {543}},\ \bibinfo {pages} {87}
  (\bibinfo {year} {2017}{\natexlab{a}})}\BibitemShut {NoStop}%
\bibitem [{\citenamefont {Dimitrova}\ \emph {et~al.}(2017)\citenamefont
  {Dimitrova}, \citenamefont {Lunden}, \citenamefont {Amato-Grill},
  \citenamefont {Jepsen}, \citenamefont {Yu}, \citenamefont {Messer},
  \citenamefont {Rigaldo}, \citenamefont {Puentes}, \citenamefont {Weld},\ and\
  \citenamefont {Ketterle}}]{Dimitrova2017}%
  \BibitemOpen
  \bibfield  {author} {\bibinfo {author} {\bibfnamefont {I.}~\bibnamefont
  {Dimitrova}}, \bibinfo {author} {\bibfnamefont {W.}~\bibnamefont {Lunden}},
  \bibinfo {author} {\bibfnamefont {J.}~\bibnamefont {Amato-Grill}}, \bibinfo
  {author} {\bibfnamefont {N.}~\bibnamefont {Jepsen}}, \bibinfo {author}
  {\bibfnamefont {Y.}~\bibnamefont {Yu}}, \bibinfo {author} {\bibfnamefont
  {M.}~\bibnamefont {Messer}}, \bibinfo {author} {\bibfnamefont
  {T.}~\bibnamefont {Rigaldo}}, \bibinfo {author} {\bibfnamefont
  {G.}~\bibnamefont {Puentes}}, \bibinfo {author} {\bibfnamefont
  {D.}~\bibnamefont {Weld}}, \ and\ \bibinfo {author} {\bibfnamefont
  {W.}~\bibnamefont {Ketterle}},\ }\href {\doibase 10.1103/PhysRevA.96.051603}
  {\bibfield  {journal} {\bibinfo  {journal} {Phys. Rev. A}\ }\textbf {\bibinfo
  {volume} {96}},\ \bibinfo {pages} {051603} (\bibinfo {year}
  {2017})}\BibitemShut {NoStop}%
\bibitem [{\citenamefont {Ostermann}\ \emph {et~al.}(2016)\citenamefont
  {Ostermann}, \citenamefont {Piazza},\ and\ \citenamefont
  {Ritsch}}]{Ostermann2016}%
  \BibitemOpen
  \bibfield  {author} {\bibinfo {author} {\bibfnamefont {S.}~\bibnamefont
  {Ostermann}}, \bibinfo {author} {\bibfnamefont {F.}~\bibnamefont {Piazza}}, \
  and\ \bibinfo {author} {\bibfnamefont {H.}~\bibnamefont {Ritsch}},\ }\href
  {\doibase 10.1103/PhysRevX.6.021026} {\bibfield  {journal} {\bibinfo
  {journal} {Phys. Rev. X}\ }\textbf {\bibinfo {volume} {6}},\ \bibinfo {pages}
  {021026} (\bibinfo {year} {2016})}\BibitemShut {NoStop}%
\bibitem [{\citenamefont {Ostermann}\ \emph {et~al.}(2017)\citenamefont
  {Ostermann}, \citenamefont {Piazza},\ and\ \citenamefont
  {Ritsch}}]{ostermann2017probing}%
  \BibitemOpen
  \bibfield  {author} {\bibinfo {author} {\bibfnamefont {S.}~\bibnamefont
  {Ostermann}}, \bibinfo {author} {\bibfnamefont {F.}~\bibnamefont {Piazza}}, \
  and\ \bibinfo {author} {\bibfnamefont {H.}~\bibnamefont {Ritsch}},\ }\href
  {\doibase 10.1088/1367-2630/aa91c3} {\bibfield  {journal} {\bibinfo
  {journal} {New J. Phys.}\ }\textbf {\bibinfo {volume} {19}},\ \bibinfo
  {pages} {125002} (\bibinfo {year} {2017})}\BibitemShut {NoStop}%
\bibitem [{\citenamefont {Kruse}\ \emph {et~al.}(2003)\citenamefont {Kruse},
  \citenamefont {Ruder}, \citenamefont {Benhelm}, \citenamefont {von Cube},
  \citenamefont {Zimmermann}, \citenamefont {Courteille}, \citenamefont
  {Els\"asser}, \citenamefont {Nagorny},\ and\ \citenamefont
  {Hemmerich}}]{Kruse2003}%
  \BibitemOpen
  \bibfield  {author} {\bibinfo {author} {\bibfnamefont {D.}~\bibnamefont
  {Kruse}}, \bibinfo {author} {\bibfnamefont {M.}~\bibnamefont {Ruder}},
  \bibinfo {author} {\bibfnamefont {J.}~\bibnamefont {Benhelm}}, \bibinfo
  {author} {\bibfnamefont {C.}~\bibnamefont {von Cube}}, \bibinfo {author}
  {\bibfnamefont {C.}~\bibnamefont {Zimmermann}}, \bibinfo {author}
  {\bibfnamefont {P.~W.}\ \bibnamefont {Courteille}}, \bibinfo {author}
  {\bibfnamefont {T.}~\bibnamefont {Els\"asser}}, \bibinfo {author}
  {\bibfnamefont {B.}~\bibnamefont {Nagorny}}, \ and\ \bibinfo {author}
  {\bibfnamefont {A.}~\bibnamefont {Hemmerich}},\ }\href {\doibase
  10.1103/PhysRevA.67.051802} {\bibfield  {journal} {\bibinfo  {journal} {Phys.
  Rev. A}\ }\textbf {\bibinfo {volume} {67}},\ \bibinfo {pages} {051802}
  (\bibinfo {year} {2003})}\BibitemShut {NoStop}%
\bibitem [{\citenamefont {Nagorny}\ \emph {et~al.}(2003)\citenamefont
  {Nagorny}, \citenamefont {Els\"asser}, \citenamefont {Richter}, \citenamefont
  {Hemmerich}, \citenamefont {Kruse}, \citenamefont {Zimmermann},\ and\
  \citenamefont {Courteille}}]{Nagorny2003}%
  \BibitemOpen
  \bibfield  {author} {\bibinfo {author} {\bibfnamefont {B.}~\bibnamefont
  {Nagorny}}, \bibinfo {author} {\bibfnamefont {T.}~\bibnamefont {Els\"asser}},
  \bibinfo {author} {\bibfnamefont {H.}~\bibnamefont {Richter}}, \bibinfo
  {author} {\bibfnamefont {A.}~\bibnamefont {Hemmerich}}, \bibinfo {author}
  {\bibfnamefont {D.}~\bibnamefont {Kruse}}, \bibinfo {author} {\bibfnamefont
  {C.}~\bibnamefont {Zimmermann}}, \ and\ \bibinfo {author} {\bibfnamefont
  {P.}~\bibnamefont {Courteille}},\ }\href {\doibase
  10.1103/PhysRevA.67.031401} {\bibfield  {journal} {\bibinfo  {journal} {Phys.
  Rev. A}\ }\textbf {\bibinfo {volume} {67}},\ \bibinfo {pages} {031401}
  (\bibinfo {year} {2003})}\BibitemShut {NoStop}%
\bibitem [{\citenamefont {Slama}\ \emph {et~al.}(2007)\citenamefont {Slama},
  \citenamefont {Krenz}, \citenamefont {Bux}, \citenamefont {Zimmermann},\ and\
  \citenamefont {Courteille}}]{Zimmermann2007}%
  \BibitemOpen
  \bibfield  {author} {\bibinfo {author} {\bibfnamefont {S.}~\bibnamefont
  {Slama}}, \bibinfo {author} {\bibfnamefont {G.}~\bibnamefont {Krenz}},
  \bibinfo {author} {\bibfnamefont {S.}~\bibnamefont {Bux}}, \bibinfo {author}
  {\bibfnamefont {C.}~\bibnamefont {Zimmermann}}, \ and\ \bibinfo {author}
  {\bibfnamefont {P.~W.}\ \bibnamefont {Courteille}},\ }\href {\doibase
  10.1103/PhysRevA.75.063620} {\bibfield  {journal} {\bibinfo  {journal} {Phys.
  Rev. A}\ }\textbf {\bibinfo {volume} {75}},\ \bibinfo {pages} {063620}
  (\bibinfo {year} {2007})}\BibitemShut {NoStop}%
\bibitem [{\citenamefont {Bux}\ \emph {et~al.}(2013)\citenamefont {Bux},
  \citenamefont {Tomczyk}, \citenamefont {Schmidt}, \citenamefont {Courteille},
  \citenamefont {Piovella},\ and\ \citenamefont {Zimmermann}}]{Bux2013}%
  \BibitemOpen
  \bibfield  {author} {\bibinfo {author} {\bibfnamefont {S.}~\bibnamefont
  {Bux}}, \bibinfo {author} {\bibfnamefont {H.}~\bibnamefont {Tomczyk}},
  \bibinfo {author} {\bibfnamefont {D.}~\bibnamefont {Schmidt}}, \bibinfo
  {author} {\bibfnamefont {P.~W.}\ \bibnamefont {Courteille}}, \bibinfo
  {author} {\bibfnamefont {N.}~\bibnamefont {Piovella}}, \ and\ \bibinfo
  {author} {\bibfnamefont {C.}~\bibnamefont {Zimmermann}},\ }\href {\doibase
  10.1103/PhysRevA.87.023607} {\bibfield  {journal} {\bibinfo  {journal} {Phys.
  Rev. A}\ }\textbf {\bibinfo {volume} {87}},\ \bibinfo {pages} {023607}
  (\bibinfo {year} {2013})}\BibitemShut {NoStop}%
\bibitem [{\citenamefont {Schmidt}\ \emph {et~al.}(2014)\citenamefont
  {Schmidt}, \citenamefont {Tomczyk}, \citenamefont {Slama},\ and\
  \citenamefont {Zimmermann}}]{Zimmermann2014}%
  \BibitemOpen
  \bibfield  {author} {\bibinfo {author} {\bibfnamefont {D.}~\bibnamefont
  {Schmidt}}, \bibinfo {author} {\bibfnamefont {H.}~\bibnamefont {Tomczyk}},
  \bibinfo {author} {\bibfnamefont {S.}~\bibnamefont {Slama}}, \ and\ \bibinfo
  {author} {\bibfnamefont {C.}~\bibnamefont {Zimmermann}},\ }\href {\doibase
  10.1103/PhysRevLett.112.115302} {\bibfield  {journal} {\bibinfo  {journal}
  {Phys. Rev. Lett.}\ }\textbf {\bibinfo {volume} {112}},\ \bibinfo {pages}
  {115302} (\bibinfo {year} {2014})}\BibitemShut {NoStop}%
\bibitem [{\citenamefont {Naik}\ \emph {et~al.}(2017)\citenamefont {Naik},
  \citenamefont {Kuyumjyan}, \citenamefont {Pandey}, \citenamefont {Bouyer},\
  and\ \citenamefont {Bertoldi}}]{Naik-ringcavitysetup}%
  \BibitemOpen
  \bibfield  {author} {\bibinfo {author} {\bibfnamefont {D.~S.}\ \bibnamefont
  {Naik}}, \bibinfo {author} {\bibfnamefont {G.}~\bibnamefont {Kuyumjyan}},
  \bibinfo {author} {\bibfnamefont {D.}~\bibnamefont {Pandey}}, \bibinfo
  {author} {\bibfnamefont {P.}~\bibnamefont {Bouyer}}, \ and\ \bibinfo {author}
  {\bibfnamefont {A.}~\bibnamefont {Bertoldi}},\ }\href
  {https://arxiv.org/abs/1712.06491} {\bibfield  {journal} {\bibinfo  {journal}
  {pre-print: arXiv:1712.06491}\ } (\bibinfo {year} {2017})}\BibitemShut
  {NoStop}%
\bibitem [{\citenamefont {Moore}\ \emph {et~al.}(1999)\citenamefont {Moore},
  \citenamefont {Zobay},\ and\ \citenamefont {Meystre}}]{Moore1990}%
  \BibitemOpen
  \bibfield  {author} {\bibinfo {author} {\bibfnamefont {M.~G.}\ \bibnamefont
  {Moore}}, \bibinfo {author} {\bibfnamefont {O.}~\bibnamefont {Zobay}}, \ and\
  \bibinfo {author} {\bibfnamefont {P.}~\bibnamefont {Meystre}},\ }\href
  {\doibase 10.1103/PhysRevA.60.1491} {\bibfield  {journal} {\bibinfo
  {journal} {Phys. Rev. A}\ }\textbf {\bibinfo {volume} {60}},\ \bibinfo
  {pages} {1491} (\bibinfo {year} {1999})}\BibitemShut {NoStop}%
\bibitem [{\citenamefont {Horak}\ and\ \citenamefont
  {Ritsch}(2001)}]{Horak2001}%
  \BibitemOpen
  \bibfield  {author} {\bibinfo {author} {\bibfnamefont {P.}~\bibnamefont
  {Horak}}\ and\ \bibinfo {author} {\bibfnamefont {H.}~\bibnamefont {Ritsch}},\
  }\href {\doibase 10.1103/PhysRevA.63.023603} {\bibfield  {journal} {\bibinfo
  {journal} {Phys. Rev. A}\ }\textbf {\bibinfo {volume} {63}},\ \bibinfo
  {pages} {023603} (\bibinfo {year} {2001})}\BibitemShut {NoStop}%
\bibitem [{\citenamefont {Nagy}\ \emph {et~al.}(2006)\citenamefont {Nagy},
  \citenamefont {Asb{\'{o}}th}, \citenamefont {Domokos},\ and\ \citenamefont
  {Ritsch}}]{Nagy2006}%
  \BibitemOpen
  \bibfield  {author} {\bibinfo {author} {\bibfnamefont {D.}~\bibnamefont
  {Nagy}}, \bibinfo {author} {\bibfnamefont {J.~K.}\ \bibnamefont
  {Asb{\'{o}}th}}, \bibinfo {author} {\bibfnamefont {P.}~\bibnamefont
  {Domokos}}, \ and\ \bibinfo {author} {\bibfnamefont {H.}~\bibnamefont
  {Ritsch}},\ }\href {\doibase 10.1209/epl/i2005-10521-4} {\bibfield  {journal}
  {\bibinfo  {journal} {Europhysics Letters ({EPL})}\ }\textbf {\bibinfo
  {volume} {74}},\ \bibinfo {pages} {254} (\bibinfo {year} {2006})}\BibitemShut
  {NoStop}%
\bibitem [{\citenamefont {Chen}\ \emph {et~al.}(2010)\citenamefont {Chen},
  \citenamefont {Goldbaum}, \citenamefont {Bhattacharya},\ and\ \citenamefont
  {Meystre}}]{Chen2010}%
  \BibitemOpen
  \bibfield  {author} {\bibinfo {author} {\bibfnamefont {W.}~\bibnamefont
  {Chen}}, \bibinfo {author} {\bibfnamefont {D.~S.}\ \bibnamefont {Goldbaum}},
  \bibinfo {author} {\bibfnamefont {M.}~\bibnamefont {Bhattacharya}}, \ and\
  \bibinfo {author} {\bibfnamefont {P.}~\bibnamefont {Meystre}},\ }\href
  {\doibase 10.1103/PhysRevA.81.053833} {\bibfield  {journal} {\bibinfo
  {journal} {Phys. Rev. A}\ }\textbf {\bibinfo {volume} {81}},\ \bibinfo
  {pages} {053833} (\bibinfo {year} {2010})}\BibitemShut {NoStop}%
\bibitem [{\citenamefont {Ostermann}\ \emph {et~al.}(2015)\citenamefont
  {Ostermann}, \citenamefont {Grie{\ss}er},\ and\ \citenamefont
  {Ritsch}}]{Ostermann2015}%
  \BibitemOpen
  \bibfield  {author} {\bibinfo {author} {\bibfnamefont {S.}~\bibnamefont
  {Ostermann}}, \bibinfo {author} {\bibfnamefont {T.}~\bibnamefont
  {Grie{\ss}er}}, \ and\ \bibinfo {author} {\bibfnamefont {H.}~\bibnamefont
  {Ritsch}},\ }\href {\doibase 10.1209/0295-5075/109/43001} {\bibfield
  {journal} {\bibinfo  {journal} {{EPL} (Europhysics Letters)}\ }\textbf
  {\bibinfo {volume} {109}},\ \bibinfo {pages} {43001} (\bibinfo {year}
  {2015})}\BibitemShut {NoStop}%
\bibitem [{\citenamefont {Ritsch}\ \emph {et~al.}(2013)\citenamefont {Ritsch},
  \citenamefont {Domokos}, \citenamefont {Brennecke},\ and\ \citenamefont
  {Esslinger}}]{Ritsch2013}%
  \BibitemOpen
  \bibfield  {author} {\bibinfo {author} {\bibfnamefont {H.}~\bibnamefont
  {Ritsch}}, \bibinfo {author} {\bibfnamefont {P.}~\bibnamefont {Domokos}},
  \bibinfo {author} {\bibfnamefont {F.}~\bibnamefont {Brennecke}}, \ and\
  \bibinfo {author} {\bibfnamefont {T.}~\bibnamefont {Esslinger}},\ }\href
  {\doibase 10.1103/RevModPhys.85.553} {\bibfield  {journal} {\bibinfo
  {journal} {Rev. Mod. Phys.}\ }\textbf {\bibinfo {volume} {85}},\ \bibinfo
  {pages} {553} (\bibinfo {year} {2013})}\BibitemShut {NoStop}%
\bibitem [{\citenamefont {Baumann}\ \emph {et~al.}(2010)\citenamefont
  {Baumann}, \citenamefont {Guerlin}, \citenamefont {Brennecke},\ and\
  \citenamefont {Esslinger}}]{Baumann2010}%
  \BibitemOpen
  \bibfield  {author} {\bibinfo {author} {\bibfnamefont {K.}~\bibnamefont
  {Baumann}}, \bibinfo {author} {\bibfnamefont {C.}~\bibnamefont {Guerlin}},
  \bibinfo {author} {\bibfnamefont {F.}~\bibnamefont {Brennecke}}, \ and\
  \bibinfo {author} {\bibfnamefont {T.}~\bibnamefont {Esslinger}},\ }\href
  {\doibase 10.1038/nature09009} {\bibfield  {journal} {\bibinfo  {journal}
  {Nature}\ }\textbf {\bibinfo {volume} {464}},\ \bibinfo {pages} {1301}
  (\bibinfo {year} {2010})}\BibitemShut {NoStop}%
\bibitem [{\citenamefont {Lang}\ \emph {et~al.}(2017)\citenamefont {Lang},
  \citenamefont {Piazza},\ and\ \citenamefont {Zwerger}}]{Lang2017}%
  \BibitemOpen
  \bibfield  {author} {\bibinfo {author} {\bibfnamefont {J.}~\bibnamefont
  {Lang}}, \bibinfo {author} {\bibfnamefont {F.}~\bibnamefont {Piazza}}, \ and\
  \bibinfo {author} {\bibfnamefont {W.}~\bibnamefont {Zwerger}},\ }\href
  {\doibase 10.1088/1367-2630/aa9b4a} {\bibfield  {journal} {\bibinfo
  {journal} {New J. Phys.}\ }\textbf {\bibinfo {volume} {19}},\ \bibinfo
  {pages} {123027} (\bibinfo {year} {2017})}\BibitemShut {NoStop}%
\bibitem [{\citenamefont {Moodie}\ \emph {et~al.}(2017)\citenamefont {Moodie},
  \citenamefont {Ballantine},\ and\ \citenamefont {Keeling}}]{Moodie2017}%
  \BibitemOpen
  \bibfield  {author} {\bibinfo {author} {\bibfnamefont {R.~I.}\ \bibnamefont
  {Moodie}}, \bibinfo {author} {\bibfnamefont {K.~E.}\ \bibnamefont
  {Ballantine}}, \ and\ \bibinfo {author} {\bibfnamefont {J.}~\bibnamefont
  {Keeling}},\ }\href {https://arxiv.org/abs/1711.03915} {\bibfield  {journal}
  {\bibinfo  {journal} {pre-print: rXiv:1711.03915}\ } (\bibinfo {year}
  {2017})}\BibitemShut {NoStop}%
\bibitem [{\citenamefont {Morales}\ \emph {et~al.}(2017)\citenamefont
  {Morales}, \citenamefont {Zupancic}, \citenamefont {Léonard}, \citenamefont
  {Esslinger},\ and\ \citenamefont {Donner}}]{Morales2017}%
  \BibitemOpen
  \bibfield  {author} {\bibinfo {author} {\bibfnamefont {A.}~\bibnamefont
  {Morales}}, \bibinfo {author} {\bibfnamefont {P.}~\bibnamefont {Zupancic}},
  \bibinfo {author} {\bibfnamefont {J.}~\bibnamefont {Léonard}}, \bibinfo
  {author} {\bibfnamefont {T.}~\bibnamefont {Esslinger}}, \ and\ \bibinfo
  {author} {\bibfnamefont {T.}~\bibnamefont {Donner}},\ }\href
  {https://arxiv.org/abs/1711.07988} {\bibfield  {journal} {\bibinfo  {journal}
  {pre-print: arXiv:1711.07988}\ } (\bibinfo {year} {2017})}\BibitemShut
  {NoStop}%
\bibitem [{\citenamefont {Grie\ss{}er}\ and\ \citenamefont
  {Ritsch}(2013)}]{Griesser2013}%
  \BibitemOpen
  \bibfield  {author} {\bibinfo {author} {\bibfnamefont {T.}~\bibnamefont
  {Grie\ss{}er}}\ and\ \bibinfo {author} {\bibfnamefont {H.}~\bibnamefont
  {Ritsch}},\ }\href {\doibase 10.1103/PhysRevLett.111.055702} {\bibfield
  {journal} {\bibinfo  {journal} {Phys. Rev. Lett.}\ }\textbf {\bibinfo
  {volume} {111}},\ \bibinfo {pages} {055702} (\bibinfo {year}
  {2013})}\BibitemShut {NoStop}%
\bibitem [{\citenamefont {Chang}\ \emph {et~al.}(2013)\citenamefont {Chang},
  \citenamefont {Cirac},\ and\ \citenamefont {Kimble}}]{chang2013}%
  \BibitemOpen
  \bibfield  {author} {\bibinfo {author} {\bibfnamefont {D.~E.}\ \bibnamefont
  {Chang}}, \bibinfo {author} {\bibfnamefont {J.~I.}\ \bibnamefont {Cirac}}, \
  and\ \bibinfo {author} {\bibfnamefont {H.~J.}\ \bibnamefont {Kimble}},\
  }\href {\doibase 10.1103/PhysRevLett.110.113606} {\bibfield  {journal}
  {\bibinfo  {journal} {Phys. Rev. Lett.}\ }\textbf {\bibinfo {volume} {110}},\
  \bibinfo {pages} {113606} (\bibinfo {year} {2013})}\BibitemShut {NoStop}%
\bibitem [{\citenamefont {Maschler}\ \emph {et~al.}(2007)\citenamefont
  {Maschler}, \citenamefont {Ritsch}, \citenamefont {Vukics},\ and\
  \citenamefont {Domokos}}]{maschler2007entanglement}%
  \BibitemOpen
  \bibfield  {author} {\bibinfo {author} {\bibfnamefont {C.}~\bibnamefont
  {Maschler}}, \bibinfo {author} {\bibfnamefont {H.}~\bibnamefont {Ritsch}},
  \bibinfo {author} {\bibfnamefont {A.}~\bibnamefont {Vukics}}, \ and\ \bibinfo
  {author} {\bibfnamefont {P.}~\bibnamefont {Domokos}},\ }\href {\doibase
  10.1016/j.optcom.2007.01.069} {\bibfield  {journal} {\bibinfo  {journal}
  {Optics Communications}\ }\textbf {\bibinfo {volume} {273}},\ \bibinfo
  {pages} {446} (\bibinfo {year} {2007})}\BibitemShut {NoStop}%
\bibitem [{\citenamefont {Vukics}\ \emph {et~al.}(2007)\citenamefont {Vukics},
  \citenamefont {Maschler},\ and\ \citenamefont
  {Ritsch}}]{vukics2007microscopic}%
  \BibitemOpen
  \bibfield  {author} {\bibinfo {author} {\bibfnamefont {A.}~\bibnamefont
  {Vukics}}, \bibinfo {author} {\bibfnamefont {C.}~\bibnamefont {Maschler}}, \
  and\ \bibinfo {author} {\bibfnamefont {H.}~\bibnamefont {Ritsch}},\ }\href
  {\doibase 10.1088/1367-2630/9/8/255} {\bibfield  {journal} {\bibinfo
  {journal} {New J. Phys.}\ }\textbf {\bibinfo {volume} {9}},\ \bibinfo {pages}
  {255} (\bibinfo {year} {2007})}\BibitemShut {NoStop}%
\bibitem [{\citenamefont {Kr\"amer}\ and\ \citenamefont
  {Ritsch}(2014)}]{kramer2014self}%
  \BibitemOpen
  \bibfield  {author} {\bibinfo {author} {\bibfnamefont {S.}~\bibnamefont
  {Kr\"amer}}\ and\ \bibinfo {author} {\bibfnamefont {H.}~\bibnamefont
  {Ritsch}},\ }\href {\doibase 10.1103/PhysRevA.90.033833} {\bibfield
  {journal} {\bibinfo  {journal} {Phys. Rev. A}\ }\textbf {\bibinfo {volume}
  {90}},\ \bibinfo {pages} {033833} (\bibinfo {year} {2014})}\BibitemShut
  {NoStop}%
\bibitem [{\citenamefont {L{\'{e}}onard}\ \emph
  {et~al.}(2017{\natexlab{b}})\citenamefont {L{\'{e}}onard}, \citenamefont
  {Morales}, \citenamefont {Zupancic}, \citenamefont {Donner},\ and\
  \citenamefont {Esslinger}}]{leonard2017monitoring}%
  \BibitemOpen
  \bibfield  {author} {\bibinfo {author} {\bibfnamefont {J.}~\bibnamefont
  {L{\'{e}}onard}}, \bibinfo {author} {\bibfnamefont {A.}~\bibnamefont
  {Morales}}, \bibinfo {author} {\bibfnamefont {P.}~\bibnamefont {Zupancic}},
  \bibinfo {author} {\bibfnamefont {T.}~\bibnamefont {Donner}}, \ and\ \bibinfo
  {author} {\bibfnamefont {T.}~\bibnamefont {Esslinger}},\ }\href {\doibase
  10.1126/science.aan2608} {\bibfield  {journal} {\bibinfo  {journal}
  {Science}\ }\textbf {\bibinfo {volume} {358}},\ \bibinfo {pages} {1415}
  (\bibinfo {year} {2017}{\natexlab{b}})}\BibitemShut {NoStop}%
\bibitem [{\citenamefont {Piazza}\ \emph {et~al.}(2013)\citenamefont {Piazza},
  \citenamefont {Strack},\ and\ \citenamefont {Zwerger}}]{piazza2013bose}%
  \BibitemOpen
  \bibfield  {author} {\bibinfo {author} {\bibfnamefont {F.}~\bibnamefont
  {Piazza}}, \bibinfo {author} {\bibfnamefont {P.}~\bibnamefont {Strack}}, \
  and\ \bibinfo {author} {\bibfnamefont {W.}~\bibnamefont {Zwerger}},\ }\href
  {\doibase http://dx.doi.org/10.1016/j.aop.2013.08.015} {\bibfield  {journal}
  {\bibinfo  {journal} {Annals of Physics}\ }\textbf {\bibinfo {volume}
  {339}},\ \bibinfo {pages} {135} (\bibinfo {year} {2013})}\BibitemShut
  {NoStop}%
\bibitem [{SM-()}]{SM-1BEC-2ringcavity}%
  \BibitemOpen
  \href@noop {} {\bibinfo  {journal} {See Supplemental Material for the details
  of the derivation of the mean-field equations, the linearized equations, and
  the threshold pump strength}\ }\BibitemShut {NoStop}%
\bibitem [{\citenamefont {Nagy}\ \emph {et~al.}(2008)\citenamefont {Nagy},
  \citenamefont {Szirmai},\ and\ \citenamefont {Domokos}}]{Nagy2008}%
  \BibitemOpen
\bibfield  {journal} {  }\bibfield  {author} {\bibinfo {author} {\bibfnamefont
  {D.}~\bibnamefont {Nagy}}, \bibinfo {author} {\bibfnamefont {G.}~\bibnamefont
  {Szirmai}}, \ and\ \bibinfo {author} {\bibfnamefont {P.}~\bibnamefont
  {Domokos}},\ }\href {\doibase 10.1140/epjd/e2008-00074-6} {\bibfield
  {journal} {\bibinfo  {journal} {Eur. Phys. J. D}\ }\textbf {\bibinfo {volume}
  {48}},\ \bibinfo {pages} {127} (\bibinfo {year} {2008})}\BibitemShut
  {NoStop}%
\bibitem [{\citenamefont {Mivehvar}\ \emph {et~al.}(2017)\citenamefont
  {Mivehvar}, \citenamefont {Piazza},\ and\ \citenamefont
  {Ritsch}}]{Mivehvar2017b}%
  \BibitemOpen
  \bibfield  {author} {\bibinfo {author} {\bibfnamefont {F.}~\bibnamefont
  {Mivehvar}}, \bibinfo {author} {\bibfnamefont {F.}~\bibnamefont {Piazza}}, \
  and\ \bibinfo {author} {\bibfnamefont {H.}~\bibnamefont {Ritsch}},\ }\href
  {\doibase 10.1103/PhysRevLett.119.063602} {\bibfield  {journal} {\bibinfo
  {journal} {Phys. Rev. Lett.}\ }\textbf {\bibinfo {volume} {119}},\ \bibinfo
  {pages} {063602} (\bibinfo {year} {2017})}\BibitemShut {NoStop}%
\bibitem [{\citenamefont {Baumann}\ \emph {et~al.}(2011)\citenamefont
  {Baumann}, \citenamefont {Mottl}, \citenamefont {Brennecke},\ and\
  \citenamefont {Esslinger}}]{Baumann2011}%
  \BibitemOpen
  \bibfield  {author} {\bibinfo {author} {\bibfnamefont {K.}~\bibnamefont
  {Baumann}}, \bibinfo {author} {\bibfnamefont {R.}~\bibnamefont {Mottl}},
  \bibinfo {author} {\bibfnamefont {F.}~\bibnamefont {Brennecke}}, \ and\
  \bibinfo {author} {\bibfnamefont {T.}~\bibnamefont {Esslinger}},\ }\href
  {\doibase 10.1103/PhysRevLett.107.140402} {\bibfield  {journal} {\bibinfo
  {journal} {Phys. Rev. Lett.}\ }\textbf {\bibinfo {volume} {107}},\ \bibinfo
  {pages} {140402} (\bibinfo {year} {2011})}\BibitemShut {NoStop}%
\bibitem [{\citenamefont {Jin}\ \emph {et~al.}(2013)\citenamefont {Jin},
  \citenamefont {Rossini}, \citenamefont {Fazio}, \citenamefont {Leib},\ and\
  \citenamefont {Hartmann}}]{Jiasen2013}%
  \BibitemOpen
  \bibfield  {author} {\bibinfo {author} {\bibfnamefont {J.}~\bibnamefont
  {Jin}}, \bibinfo {author} {\bibfnamefont {D.}~\bibnamefont {Rossini}},
  \bibinfo {author} {\bibfnamefont {R.}~\bibnamefont {Fazio}}, \bibinfo
  {author} {\bibfnamefont {M.}~\bibnamefont {Leib}}, \ and\ \bibinfo {author}
  {\bibfnamefont {M.~J.}\ \bibnamefont {Hartmann}},\ }\href {\doibase
  10.1103/PhysRevLett.110.163605} {\bibfield  {journal} {\bibinfo  {journal}
  {Phys. Rev. Lett.}\ }\textbf {\bibinfo {volume} {110}},\ \bibinfo {pages}
  {163605} (\bibinfo {year} {2013})}\BibitemShut {NoStop}%
\bibitem [{\citenamefont {Bujnowski}\ \emph {et~al.}(2014)\citenamefont
  {Bujnowski}, \citenamefont {Corso}, \citenamefont {Hayward}, \citenamefont
  {Cole},\ and\ \citenamefont {Martin}}]{Bujnowski2014}%
  \BibitemOpen
  \bibfield  {author} {\bibinfo {author} {\bibfnamefont {B.}~\bibnamefont
  {Bujnowski}}, \bibinfo {author} {\bibfnamefont {J.~K.}\ \bibnamefont
  {Corso}}, \bibinfo {author} {\bibfnamefont {A.~L.~C.}\ \bibnamefont
  {Hayward}}, \bibinfo {author} {\bibfnamefont {J.~H.}\ \bibnamefont {Cole}}, \
  and\ \bibinfo {author} {\bibfnamefont {A.~M.}\ \bibnamefont {Martin}},\
  }\href {\doibase 10.1103/PhysRevA.90.043801} {\bibfield  {journal} {\bibinfo
  {journal} {Phys. Rev. A}\ }\textbf {\bibinfo {volume} {90}},\ \bibinfo
  {pages} {043801} (\bibinfo {year} {2014})}\BibitemShut {NoStop}%
\bibitem [{\citenamefont {Landig}\ \emph {et~al.}(2016)\citenamefont {Landig},
  \citenamefont {Hruby}, \citenamefont {Dogra}, \citenamefont {Landini},
  \citenamefont {Mottl}, \citenamefont {Donner},\ and\ \citenamefont
  {Esslinger}}]{Landig2016}%
  \BibitemOpen
  \bibfield  {author} {\bibinfo {author} {\bibfnamefont {R.}~\bibnamefont
  {Landig}}, \bibinfo {author} {\bibfnamefont {L.}~\bibnamefont {Hruby}},
  \bibinfo {author} {\bibfnamefont {N.}~\bibnamefont {Dogra}}, \bibinfo
  {author} {\bibfnamefont {M.}~\bibnamefont {Landini}}, \bibinfo {author}
  {\bibfnamefont {R.}~\bibnamefont {Mottl}}, \bibinfo {author} {\bibfnamefont
  {T.}~\bibnamefont {Donner}}, \ and\ \bibinfo {author} {\bibfnamefont
  {T.}~\bibnamefont {Esslinger}},\ }\href {\doibase 10.1038/nature17409}
  {\bibfield  {journal} {\bibinfo  {journal} {Nature}\ }\textbf {\bibinfo
  {volume} {532}},\ \bibinfo {pages} {476} (\bibinfo {year}
  {2016})}\BibitemShut {NoStop}%
\bibitem [{\citenamefont {Berman}(1997)}]{berman1997atom}%
  \BibitemOpen
  \bibfield  {author} {\bibinfo {author} {\bibfnamefont {P.~R.}\ \bibnamefont
  {Berman}},\ }\href@noop {} {\emph {\bibinfo {title} {Atom interferometry}}}\
  (\bibinfo  {publisher} {Academic press},\ \bibinfo {year} {1997})\BibitemShut
  {NoStop}%
\bibitem [{\citenamefont {Cronin}\ \emph {et~al.}(2009)\citenamefont {Cronin},
  \citenamefont {Schmiedmayer},\ and\ \citenamefont {Pritchard}}]{cronin_2009}%
  \BibitemOpen
  \bibfield  {author} {\bibinfo {author} {\bibfnamefont {A.~D.}\ \bibnamefont
  {Cronin}}, \bibinfo {author} {\bibfnamefont {J.}~\bibnamefont
  {Schmiedmayer}}, \ and\ \bibinfo {author} {\bibfnamefont {D.~E.}\
  \bibnamefont {Pritchard}},\ }\href {\doibase 10.1103/RevModPhys.81.1051}
  {\bibfield  {journal} {\bibinfo  {journal} {Rev. Mod. Phys.}\ }\textbf
  {\bibinfo {volume} {81}},\ \bibinfo {pages} {1051} (\bibinfo {year}
  {2009})}\BibitemShut {NoStop}%
\bibitem [{\citenamefont {Sorrentino}\ \emph {et~al.}(2010)\citenamefont
  {Sorrentino}, \citenamefont {Bongs}, \citenamefont {Bouyer}, \citenamefont
  {Cacciapuoti}, \citenamefont {de~Angelis}, \citenamefont {Dittus},
  \citenamefont {Ertmer}, \citenamefont {Giorgini}, \citenamefont {Hartwig},
  \citenamefont {Hauth}, \citenamefont {Herrmann}, \citenamefont {Inguscio},
  \citenamefont {Kajari}, \citenamefont {Könemann}, \citenamefont
  {Lämmerzahl}, \citenamefont {Landragin}, \citenamefont {Modugno},
  \citenamefont {Pereira~dos Santos}, \citenamefont {Peters}, \citenamefont
  {Prevedelli}, \citenamefont {Rasel}, \citenamefont {Schleich}, \citenamefont
  {Schmidt}, \citenamefont {Senger}, \citenamefont {Sengstock}, \citenamefont
  {Stern}, \citenamefont {Tino},\ and\ \citenamefont {Walser}}]{tinoetal_2010}%
  \BibitemOpen
  \bibfield  {author} {\bibinfo {author} {\bibfnamefont {F.}~\bibnamefont
  {Sorrentino}}, \bibinfo {author} {\bibfnamefont {K.}~\bibnamefont {Bongs}},
  \bibinfo {author} {\bibfnamefont {P.}~\bibnamefont {Bouyer}}, \bibinfo
  {author} {\bibfnamefont {L.}~\bibnamefont {Cacciapuoti}}, \bibinfo {author}
  {\bibfnamefont {M.}~\bibnamefont {de~Angelis}}, \bibinfo {author}
  {\bibfnamefont {H.}~\bibnamefont {Dittus}}, \bibinfo {author} {\bibfnamefont
  {W.}~\bibnamefont {Ertmer}}, \bibinfo {author} {\bibfnamefont
  {A.}~\bibnamefont {Giorgini}}, \bibinfo {author} {\bibfnamefont
  {J.}~\bibnamefont {Hartwig}}, \bibinfo {author} {\bibfnamefont
  {M.}~\bibnamefont {Hauth}}, \bibinfo {author} {\bibfnamefont
  {S.}~\bibnamefont {Herrmann}}, \bibinfo {author} {\bibfnamefont
  {M.}~\bibnamefont {Inguscio}}, \bibinfo {author} {\bibfnamefont
  {E.}~\bibnamefont {Kajari}}, \bibinfo {author} {\bibfnamefont
  {T.}~\bibnamefont {Könemann}}, \bibinfo {author} {\bibfnamefont
  {C.}~\bibnamefont {Lämmerzahl}}, \bibinfo {author} {\bibfnamefont
  {A.}~\bibnamefont {Landragin}}, \bibinfo {author} {\bibfnamefont
  {G.}~\bibnamefont {Modugno}}, \bibinfo {author} {\bibfnamefont
  {F.}~\bibnamefont {Pereira~dos Santos}}, \bibinfo {author} {\bibfnamefont
  {A.}~\bibnamefont {Peters}}, \bibinfo {author} {\bibfnamefont
  {M.}~\bibnamefont {Prevedelli}}, \bibinfo {author} {\bibfnamefont
  {E.}~\bibnamefont {Rasel}}, \bibinfo {author} {\bibfnamefont
  {W.}~\bibnamefont {Schleich}}, \bibinfo {author} {\bibfnamefont
  {M.}~\bibnamefont {Schmidt}}, \bibinfo {author} {\bibfnamefont
  {A.}~\bibnamefont {Senger}}, \bibinfo {author} {\bibfnamefont
  {K.}~\bibnamefont {Sengstock}}, \bibinfo {author} {\bibfnamefont
  {G.}~\bibnamefont {Stern}}, \bibinfo {author} {\bibfnamefont
  {G.}~\bibnamefont {Tino}}, \ and\ \bibinfo {author} {\bibfnamefont
  {R.}~\bibnamefont {Walser}},\ }\href {\doibase 10.1007/s12217-010-9240-7}
  {\bibfield  {journal} {\bibinfo  {journal} {Microgravity Science and
  Technology}\ }\textbf {\bibinfo {volume} {22}},\ \bibinfo {pages} {551}
  (\bibinfo {year} {2010})}\BibitemShut {NoStop}%
\bibitem [{\citenamefont {Hamilton}\ \emph {et~al.}(2015)\citenamefont
  {Hamilton}, \citenamefont {Jaffe}, \citenamefont {Brown}, \citenamefont
  {Maisenbacher}, \citenamefont {Estey},\ and\ \citenamefont
  {M\"uller}}]{mueller_atomint_cavity_2015}%
  \BibitemOpen
  \bibfield  {author} {\bibinfo {author} {\bibfnamefont {P.}~\bibnamefont
  {Hamilton}}, \bibinfo {author} {\bibfnamefont {M.}~\bibnamefont {Jaffe}},
  \bibinfo {author} {\bibfnamefont {J.~M.}\ \bibnamefont {Brown}}, \bibinfo
  {author} {\bibfnamefont {L.}~\bibnamefont {Maisenbacher}}, \bibinfo {author}
  {\bibfnamefont {B.}~\bibnamefont {Estey}}, \ and\ \bibinfo {author}
  {\bibfnamefont {H.}~\bibnamefont {M\"uller}},\ }\href {\doibase
  10.1103/PhysRevLett.114.100405} {\bibfield  {journal} {\bibinfo  {journal}
  {Phys. Rev. Lett.}\ }\textbf {\bibinfo {volume} {114}},\ \bibinfo {pages}
  {100405} (\bibinfo {year} {2015})}\BibitemShut {NoStop}%
\end{thebibliography}%

%=========================================================================
\newpage
\widetext
\setcounter{equation}{0}
\renewcommand{\theequation}{S\arabic{equation}}

%=========================================================================
\section{Supplemental Material}

Here we present the details of the derivation of the mean-field equations [Eq.~(2) in the main text], 
the linearized equations [Eqs.~(3) and (4) in the main text], and the threshold pump strength [Eq.~(5) in the main text].

%=========================================================================
\section{Mean-Field Equations}

The Heisenberg equations of motion of the 
photonic and atomic field operators can be obtained using  
the many-body effective Hamiltonian~$H_{\rm eff}$, given in the manuscript, as
\begin{align} \label{eqSM:Heisenberg-eqs}
i\hbar\frac{\partial}{\partial t}\hat{a}_\pm&=[\hat{a}_\pm,H_{\rm eff}]=
\hbar\left(-\Delta_c+U\hat{N}-i\kappa\right)\hat{a}_\pm
+\hbar U\hat{\mathcal N}_{\pm2}\hat{a}_\mp
+\hbar\eta\hat{\mathcal N}_{\pm1},
\nonumber\\
i\hbar\frac{\partial}{\partial t}\hat\psi&=[\hat\psi,H_{\rm eff}]=
\mathcal{H}_{\rm eff}^{(1)}\hat\psi,
\end{align}
where $\hat{N}=\int \hat\psi^\dag(x)\hat\psi(x) dx$,
$\hat{\mathcal N}_{\pm1}\equiv\int \hat\psi^\dag(x) e^{\mp ik_cx} \hat\psi(x)dx$, and
$\hat{\mathcal N}_{\pm2}\equiv\int \hat\psi^\dag(x) e^{\mp2ik_cx} \hat\psi(x)dx$.
Here we have added field damping terms proportional to the cavity  decay rate $\kappa$.
By replace the photonic and atomic field operators with their corresponding quantum averages,
$\hat{a}_\pm(t)\to \langle\hat{a}_\pm(t)\rangle=\alpha_\pm(t)=|\alpha_\pm(t)|e^{i\phi_\pm(t)}$ and 
$\hat\psi(x,t)\to\langle \hat\psi(x,t)\rangle=\psi(x,t)=\sqrt{n(x,t)}e^{i\theta(t)}$, respectively,
one obtains the three mean-field coupled equations (2) in the manuscript.

%=========================================================================
\section{Linearized Equations}

Assuming $\psi(x,t)=e^{-i\mu_0 t/\hbar}[\psi_0(x)+\delta\psi(x,t)]$ and 
$\alpha_\pm(t)=\alpha_{0\pm}+\delta\alpha_\pm(t)$, where 
$\psi_0(x)$ and $\alpha_{0\pm}$ are the mean-field stationary-state solutions
of Eq.~(2) in the main text with the chemical potential $\mu_0$, 
linearizing Eq.~(2) yields
\begin{align} \label{eqSM:linearized-eqs}
i\frac{\partial}{\partial t}\delta\alpha_\pm&=
\left(-\Delta_c+UN-i\kappa\right)\delta\alpha_\pm
+U\mathcal{N}_{\pm2}^{(0)}\delta\alpha_\mp
+\int A_{\pm}(\psi_0^*\delta\psi+\psi_0\delta\psi^*)dx,
\nonumber\\
i\frac{\partial}{\partial t}\delta\psi&=
\frac{1}{\hbar}\left(\mathcal{H}_{\rm eff}^{(1)}-\mu_0\right)\delta\psi
+\psi_0 \left(A_+^*\delta\alpha_+ + A_+\delta\alpha_+^* + A_-^*\delta\alpha_- + A_-\delta\alpha_-^* \right),
\end{align}
where $\mathcal{N}_{\pm2}^{(0)}=\int n_0(x)e^{\mp2ik_cx}dx$ and we have defined 
$A_{\pm}(x)\equiv U \left(\alpha_{0\pm}+\alpha_{0\mp} e^{\mp2ik_cx}\right)+\eta e^{\mp ik_cx}$ for shorthands. 
Since the linearized equations~\eqref{eqSM:linearized-eqs} couple $\delta\psi$ and $\delta\alpha_\pm$ to their complex conjugates,
we make the ans\"{a}tze 
$\delta\psi(x,t)=\delta\psi^{(+)}(x)e^{-i\omega t}
+[\delta\psi^{(-)}(x)]^*e^{i\omega^* t}$
and $\delta\alpha_\pm(t)=\delta\alpha_\pm^{(+)}e^{-i\omega t}
+[\delta\alpha_\pm^{(-)}]^*e^{i\omega^* t}$ for the quantum fluctuations. 
Substituting these ans\"{a}tze 
in Eq.~\eqref{eqSM:linearized-eqs} and setting the coefficients of $e^{-i\omega t}$
and $e^{i\omega^*t}$ separately to zero 
yields a set of six coupled Bogoliubov-type equations for the positive-
and negative-frequency components of the quantum fluctuations,  
\begin{align} \label{eqSM:linearized-eqs-pos-neg}
\omega\delta\alpha_\pm^{(+)}&=
\left(-\Delta_c+UN-i\kappa\right)\delta\alpha_\pm^{(+)}
+U\mathcal{N}_{\pm2}^{(0)}\delta\alpha_\mp^{(+)}
+\int A_{\pm}\left[\psi_0^*\delta\psi^{(+)}+\psi_0\delta\psi^{(-)}\right]dx,
\nonumber\\
\omega\delta\alpha_\pm^{(-)}&=
-\left(-\Delta_c+UN-i\kappa\right)^*\delta\alpha_\pm^{(-)}
-U\mathcal{N}_{\pm2}^{(0)*}\delta\alpha_\mp^{(-)}
-\int A_{\pm}^*\left[\psi_0^*\delta\psi^{(+)}+\psi_0\delta\psi^{(-)}\right]dx,
\nonumber\\
\omega\delta\psi^{(+)}&=
\frac{1}{\hbar}\left[\mathcal{H}_{\rm eff}^{(1)}-\mu_0\right]\delta\psi^{(+)}
+\psi_0 
\left[A_+^*\delta\alpha_+^{(+)} + A_+\delta\alpha_+^{(-)} + A_-^*\delta\alpha_-^{(+)} + A_-\delta\alpha_-^{(-)}\right],
\nonumber\\
\omega\delta\psi^{(-)}&=
-\frac{1}{\hbar}\left[\mathcal{H}_{\rm eff}^{(1)}-\mu_0\right]\delta\psi^{(-)}
-\psi_0^* 
\left[A_+^*\delta\alpha_+^{(+)} + A_+\delta\alpha_+^{(-)} + A_-^*\delta\alpha_-^{(+)} + A_-\delta\alpha_-^{(-)}\right].
\end{align}
We recast these equations in a matrix form 
%$\omega\mathbf{f}=\mathbf{M}_{\rm B}\mathbf{f}$,
\begin{align} \label{eqSM:Bog-eq}
\omega\mathbf{f}=\mathbf{M}_{\rm B}\mathbf{f},
\end{align}
where 
$\mathbf{f}=(
\delta\alpha_+^{(+)},\delta\alpha_+^{(-)},
\delta\alpha_-^{(+)},\delta\alpha_-^{(-)},
\delta\psi^{(+)},\delta\psi^{(-)})^\mathsf{T}$
and 
\begin{align} \label{eqSM:Bog-matrix}
\renewcommand*{\arraystretch}{1.25}
\mathbf{M}_{\rm B}=
\begin{pmatrix}
\delta_c & 0 & U\mathcal{N}_{+2}^{(0)} & 0 & \mathcal{I}_{+*} & \mathcal{I}_{+} \\
0 & -\delta_c^* & 0 & -U\mathcal{N}_{+2}^{(0)*} & -\mathcal{I}_{+}^* & -\mathcal{I}_{+*}^*  \\
U\mathcal{N}_{-2}^{(0)} & 0 & \delta_c & 0 & \mathcal{I}_{-*} & \mathcal{I}_{-} \\
0 & -U\mathcal{N}_{-2}^{(0)*} & 0 & -\delta_c^* & -\mathcal{I}_{-}^* & -\mathcal{I}_{-*}^*\\
\psi_0A_+^* & \psi_0A_+ & \psi_0A_-^* & \psi_0A_- & (\mathcal{H}_{\rm eff}^{(1)}-\mu_0)/\hbar & 0 \\
-\psi_0^*A_+^* & -\psi_0^*A_+ & -\psi_0^*A_-^* & -\psi_0^*A_- & 0 & -(\mathcal{H}_{\rm eff}^{(1)}-\mu_0)/\hbar
\end{pmatrix},
\end{align}
with $\delta_c\equiv-\Delta_c+UN-i\kappa$.
Here we have introduced the integral operators,
\begin{align}
\mathcal{I}_{\pm}\xi&=\int A_\pm(x) \psi_0(x)\xi dx,
\nonumber\\
\mathcal{I}_{\pm*}\xi&=\int A_\pm(x) \psi_0^*(x)\xi dx.
%\nonumber\\
%\mathcal{I}_{\pm*}^{(-)}\xi&=\int A_\pm^*(x) \psi_0^*(x)\xi dx,
%\nonumber\\
%\mathcal{I}_{\pm}^{(-)}\xi&=\int A_\pm^*(x) \psi_0(x)\xi dx.
\end{align}
We find eigenvalues $\omega$ of Eq.~\eqref{eqSM:Bog-eq} by numerically diagonalizing 
the Bogoliubov matrix~\eqref{eqSM:Bog-matrix} on one unit cell (of length $\lambda_c$)
with a periodic boundary condition. We discretize the space and replace the
kinetic energy term $-(\hbar^2/2m)\partial_x^2$ and the integral operators 
$\{\mathcal{I}_\pm,\mathcal{I}_{\pm*}\}$ with the corresponding
finite-difference terms.

%--------------------------------------------------------------------------------------
\subsection{The Threshold Pump Strength}

In order to find an analytical equation for the critical threshold, we analyze the stability
of the the trivial solution (i.e., the solution below the Dicke transition) 
$\alpha_{0\pm}=0$ and $\psi_0=\sqrt{N/\lambda_c}$ with $\mu_0=0$, 
by restricting the atomic fluctuations to the momentum states $\pm\hbar k_c$.
Using this trivial solution and the ansatz 
$\delta\psi^{(\pm)}(x)=\delta\psi_+^{(\pm)}e^{ik_cx}+\delta\psi_-^{(\pm)}e^{-ik_cx}$ 
for the positive- and negative-frequency condensate fluctuations with momenta $\pm\hbar k_c$,   
the Bogoliubov matrix~\eqref{eqSM:Bog-matrix} takes the following form, 
\begin{align} \label{eqSM:Bog-matrix-triv-sol}
\tilde{\mathbf{M}}_{\rm B}=
\begin{pmatrix}
\delta_c & 0 & 0 & 0 & \sqrt{N\lambda_c}\eta & \sqrt{N\lambda_c}\eta & 0 & 0 \\
0 & -\delta_c^* & 0 & 0 & 0 & 0 & -\sqrt{N\lambda_c}\eta & -\sqrt{N\lambda_c}\eta  \\
0 & 0 & \delta_c & 0 & 0 & 0& \sqrt{N\lambda_c}\eta & \sqrt{N\lambda_c}\eta \\
0 & 0 & 0 & -\delta_c^* & -\sqrt{N\lambda_c}\eta & -\sqrt{N\lambda_c}\eta & 0 & 0 \\
\sqrt{N/\lambda_c}\eta & 0 & 0 & \sqrt{N/\lambda_c}\eta & \omega_r & 0 & 0 & 0 \\
-\sqrt{N/\lambda_c}\eta & 0 & 0 & -\sqrt{N/\lambda_c}\eta & 0 & -\omega_r & 0 & 0 \\
0 & \sqrt{N/\lambda_c}\eta & \sqrt{N/\lambda_c}\eta & 0 & 0 & 0 & \omega_r & 0 \\
0 & -\sqrt{N/\lambda_c}\eta & -\sqrt{N/\lambda_c}\eta & 0 & 0 & 0 & 0 & -\omega_r
\end{pmatrix}.
\end{align}
The eigenvalues $\omega$ of $\tilde{\mathbf{M}}_{\rm B}$ is obtained via the
eighth-order characteristic equation $\text{Det}(\tilde{\mathbf{M}}_{\rm B}-\omega I_{8\times8})=0$:  
\begin{align} \label{eqSM:charact-eq}
[(\omega^2-\omega_r^2)(\omega-\delta_c)(\omega+\delta_c^*)+4N\eta^2\omega_r(\Delta_c-UN)]^2=0.
\end{align}
The solution of the characteristic equation~\eqref{eqSM:charact-eq}
yields the spectra $\omega$ of the atomic and photonic excitations, which below the threshold $\eta_c$ 
are in excellent agreement with the first and last excitation bands of Fig.~3 
in the manuscript obtained from the full numerical calculations. 
Above $\eta_c$ the solutions of Eq.~\eqref{eqSM:charact-eq}
develop positive imaginary parts, signaling that the trivial solution
$\alpha_{0\pm}=0$ and $\psi_0=\sqrt{N/\lambda_c}$ is 
unstable towards the superradiant phase. 
The zero-frequency solution $\omega=0$ of the characteristic equation~\eqref{eqSM:charact-eq} 
yields the self-ordering threshold,
\begin{align}
\sqrt{N}\eta_c=\sqrt{\frac{(-\Delta_c+UN)^2+\kappa^2}{4(-\Delta_c+UN)}}\sqrt{\omega_r}.
\end{align}

\end{document}